\documentclass[draftclsnofoot,onecolumn,12pt]{IEEEtran}

\usepackage{cite}
\usepackage{hyperref}
\ifCLASSINFOpdf
\else
\usepackage[dvips]{graphicx}
\fi

\usepackage{url}

\usepackage{amssymb,amsmath,amsthm}
\usepackage{array}

\usepackage[small]{caption}
\usepackage{subcaption}

\usepackage{multirow}
\usepackage{psfrag}
\usepackage{cleveref}

\usepackage{xspace}

\newtheorem{theorem}{Theorem}
\newtheorem{lemma}{Lemma}
\newtheorem{fact}{Fact}
\newtheorem{proposition}{Proposition}
\newtheorem{corollary}{Corollary}
\newtheorem{definition}{Definition}
\DeclareMathOperator*{\argmax}{arg\,max}

\DeclareMathOperator*{\sinc}{sinc}

\DeclareMathOperator*{\erfc}{erfc}
\renewcommand{\P}{\mathbb{P}}
\newcommand{\R}{\mathbb{R}}
\newcommand{\E}{\mathbb{E}}
\newcommand{\ie}{\emph{i.e.,}\@\xspace}
\newcommand{\eg}{\emph{e.g.,}\@\xspace}
\renewcommand{\L}{\mathcal{L}}
\renewcommand{\d}{\textnormal{d}}
\def\peq#1{\stackrel{\text{\scriptsize(#1)}}{=}}
\def\pgeq#1{\stackrel{\text{\scriptsize(#1)}}{\geq}}
\def\pleq#1{\stackrel{\text{\scriptsize(#1)}}{\leq}}

\begin{document}
\title{The Performance of Successive Interference Cancellation in Random Wireless Networks}

\author{Xinchen~Zhang
and~Martin~Haenggi  \vspace{-0.8cm}
\thanks{Manuscript date \today. The corresponding author is Xinchen Zhang ({\tt x.zhang@utexas.edu}).
Part of this paper was presented in 2012 IEEE Global Communications Conference (GLOBECOM'12) and 2013 IEEE International Symposium on Information Theory (ISIT'13).
This work was partially supported by the NSF (grants CNS 1016742
and CCF 1216407).}}%
\markboth{}{Updated \today}%

\maketitle

\begin{abstract}

This paper provides a unified framework to study the performance
of successive interference cancellation (SIC) in 
wireless networks with arbitrary fading distribution
and power-law path loss.
An analytical characterization of the performance of SIC
is given as a function of different system parameters. 
The results suggest that the marginal benefit of enabling the receiver to successively decode $k$ users
diminishes very fast with $k$,
especially in networks of high dimensions and small path loss exponent.
On the other hand, SIC is highly beneficial when the users are clustered around
the receiver and/or very low-rate codes are used.
Also, with multiple packet reception,
a lower per-user information rate always results in higher
aggregate throughput in interference-limited networks.
In contrast, there exists
a positive optimal per-user rate that maximizes the aggregate throughput in noisy networks.

The analytical results serve as useful tools to understand the potential gain
of SIC in heterogeneous cellular networks (HCNs).
Using these tools, this paper quantifies the gain of SIC on the coverage probability
in HCNs with non-accessible base stations.
An interesting observation is that, for contemporary narrow-band systems (\emph{e.g.,} LTE and WiFi),
most of the gain of SIC is achieved by canceling a single interferer.

\end{abstract}


\section{Introduction}

Although suboptimal in general,
successive interference cancellation (SIC) is a promising
technique to improve the efficiency of the wireless networks
with relatively small additional complexity
\cite{net:Weber07tit,BaccelliGamalTse11tit}.
However, in a network without centralized power control,
\emph{e.g.,} ad hoc networks,
the use of SIC hinges on the imbalance of the received powers
from different users (active transmitters),
which depends on the spatial distribution of the users
as well as many other network parameters.
Therefore, it is important to quantify the gain of SIC
with respect to different system parameters.

This paper provides a unified framework to study the performance of SIC
in $d$-dimensional wireless networks.
Modeling the active transmitters in the network by
a Poisson point process (PPP) with power-law density function
(which includes the uniform PPP as a special case),
we show how the effectiveness of SIC depends on the path loss exponent,
fading, coding rate, and user distribution.
As an application of the technical results,
we study the performance of SIC in heterogeneous cellular networks (HCNs) in the end of the paper.

\subsection{Successive Interference Cancellation and Related Work}

As contemporary wireless systems are becoming increasingly interference-limited,
there is an ascending interest in using advanced interference mitigation techniques to
improve the network performance in addition to the conventional approach of treating interference as background noise
\cite{net:Weber07tit,net:Hunter08twc,BlomerJindal09icc,BaccelliGamalTse11tit,CadambeJafar09tit,Costa83,MiridakisVergados12,HuangAndrewsGuoHeathBerry12tit}.
One important approach is successive interference cancellation (SIC).
First introduced in \cite{Cover72},
the idea of SIC is to decode different users sequentially,
\ie the interference due to the decoded users is subtracted before decoding other users.
Although SIC is not always the optimal multiple access scheme in wireless networks \cite{BlomerJindal09icc,BaccelliGamalTse11tit},
it is especially amenable to implementation \cite{Andrews05,net:Vanka12twc,HalperinWetherall2008Mobicom_Sting}
and does attain boundaries of the capacity regions in multiuser systems in many cases\cite{BaccelliGamalTse11tit,net:Cover91book,RimoldiUrbanke96}.

Conventional performance analyses of SIC do not take into account the spatial distribution of the users.
The transmitters are either assumed to reside at given locations with deterministic path loss,
see, \emph{e.g.,} \cite{ZanellaZorzi12} and the references therein,
or assumed subject to centralized power control which to a large extent compensates for the channel randomness \cite{net:Viterbi90,AndrewsMeng03}.
To establish advanced models that take into account the spatial distribution of the users,
recent papers attempt to analyze the performance of SIC
using tools from stochastic geometry \cite{net:Haenggi09jsac,net:mh12}.
In this context, 
a \emph{guard-zone} based approximation is often used to model the effect of interference cancellation
due to the well-acknowledged difficulty in tackling the problem directly \cite{net:Weber07tit}.
According to this approximation, the interferers inside a guard-zone centered at the receiver are
assumed canceled, and the size of the guard-zone is used to model the SIC capability.
Despite many interesting results obtained by this approximation,
it does not provide enough insights on the effect of received power ordering from
different transmitters, which is essential for successive decoding\cite{net:Viterbi90}. 
For example, if there are two or more (active) transmitters at the same distance to the receiver,
it is very likely that none of them can be decoded given the fact that the decoding
requires a reasonable SINR, \eg no less than one, 
while the guard-zone model would assume they all can be decoded if they are in the guard zone.
Therefore, the guard-zone approach provides a good approximation only for canceling one or at most two interferers.
Furthermore, most of the work in this line of research considers Rayleigh fading and/or uniformly
distributed networks.
In contrast, this paper uses an exact approach to tackle
the problem directly for a more general type (non-uniform) of networks with arbitrary fading distribution.

Besides SIC, there are many other techniques that can potentially significantly mitigate the interference in
wireless networks including interference alignment \cite{CadambeJafar09tit} and dirty paper coding \cite{Costa83}.
Despite the huge promise in terms of performance gain, these techniques typically rely heavily on
accurate channel state information at the transmitters (CSIT) and thus are less likely to
impact practical wireless systems in the near future \cite{MiridakisVergados12,HuangAndrewsGuoHeathBerry12tit}.
Also, many recent works study interference cancellation based on MIMO techniques in the context of random wireless networks,
\emph{e.g.,} \cite{HuangAndrewsGuoHeathBerry12tit,VazeHeath12} and references therein.
These (linear) interference cancellation techniques should not be considered as successive interference cancellation (SIC),
although they can be combined with SIC to achieve (even) better performance\cite{tse2005fundamentals}.

\subsection{Contributions and Organization}

This paper considers SIC as a pure receiver end technique\footnote{In general, SIC can be combined with (centralized) power
control, which can significantly boost its usefulness. However, this places extra overhead in transmitter coordination
and is beyond the discussion of this paper.}, which does not require any modifications to the conventional
transmitter architecture.
With a general framework for the analysis of $d$-dimensional Poisson networks,
the primary focus of this paper is on 2-d networks\footnote{Although the most interesting case is the planar networks ($d=2$)
and it may be helpful to always think of the 2-d case while reading this paper,
it is worth noting that the case $d=1$ is also of interest as it has natural applications in vehicular networks.
}, where all the nodes are transmitting at the same rate.

The main contributions of this paper are summarized as follows:

\begin{itemize}
\item
We show that fading does not affect the performance of SIC in 
a large class of interference-limited networks, including uniform networks as a special case (Section~\ref{sec:PLPF}).
However, in noisy networks, fading always reduces the decoding probability (Section~\ref{sec:noise}).

\item
We provide a set of 
closed-form upper and lower bounds on the probability of successively decoding at
least $k$ users.
These bounds are based on different ideas and are reasonably tight in different regimes (Section~\ref{sec:boundsonpk}).

\item
In interference-limited networks,
when the per-user information rate goes to 0,
we show that the aggregate throughput at the receiver is upper bounded by $\frac{1}{\beta}-1$,
where $\beta$ is a simple function of the path loss exponent, network density and network dimensionality.
A Laplace transform-based approximation is also found for the aggregate throughput at the receiver
for general per user information rate (Section~\ref{sec:AggThroughput}).

\item
We observe that in interference-limited network the aggregate throughput at a typical receiver
is a monotonically decreasing function of the per user information rate,
while in noisy networks (Section~\ref{sec:AggThroughput}),
there exists an optimal positive per-user rate that maximizes
the aggregate throughput (Section~\ref{sec:noise}).

\item
We provide an example to illustrate how the results of this paper can be applied to heterogeneous
cellular networks (HCNs).
The results demonstrate that SIC can boost the coverage probability in heterogeneous networks
with overloaded or closed-access base stations (Section~\ref{sec:HetNet}).
However, SIC is not very helpful in terms of average throughput for typical system parameters.
Moreover, for typical contemporary OFDM-based systems,
most of the gain of SIC comes from canceling a single interferer (Section~\ref{subsec:FiniteSIC}).

\end{itemize}

The rest of the paper is organized as follows:
Section~\ref{sec:sysmodel} describes the system models and the metrics
we are using in this paper.
Section~\ref{sec:PLPF} introduces the path loss process with fading (PLPF)-based (narrow band)
framework which facilitates the analysis in the rest of the paper.
In Section~\ref{sec:boundsonpk}, we provide a set of bounds on
the probability of decoding at least $k$ users in system.
These bounds directly lead to bounds on the expected gain of SIC presented
in Section~\ref{sec:EN}.
We discuss the effect of noise in Section~\ref{sec:noise}.
Section~\ref{sec:HetNet} applies the results to the downlink of HCNs.
The paper is concluded in Section~\ref{sec:conclu}.

\section{System Model and Metrics\label{sec:sysmodel}}

\subsection{The Power-law Poisson Network with Fading (PPNF)}

Let the receiver be at the origin $o$ and the active transmitters (users)
be represented by a marked Poisson point process (PPP)
$\hat{\Phi}=\{(x_i,h_{x_i})\}\subset \mathbb{R}^d \times \mathbb{R}^{+}$,
where $x$ is the location of a user,
$h_x$ is the iid (power) fading coefficient associated with the link
from $x$ to $o$,
and $d$ is the number of dimensions of the space.
When the ground process $\Phi\subset \R^d$ is a homogeneous PPP,
the network is termed a \emph{homogeneous Poisson network}
which is often the focus of stochastic geometry-based network analyses.

\begin{figure}[t]
\centering
{
	\begin{minipage}[b]{.28\linewidth}
		\centering
		{
		\includegraphics[width=\linewidth,height=\linewidth]{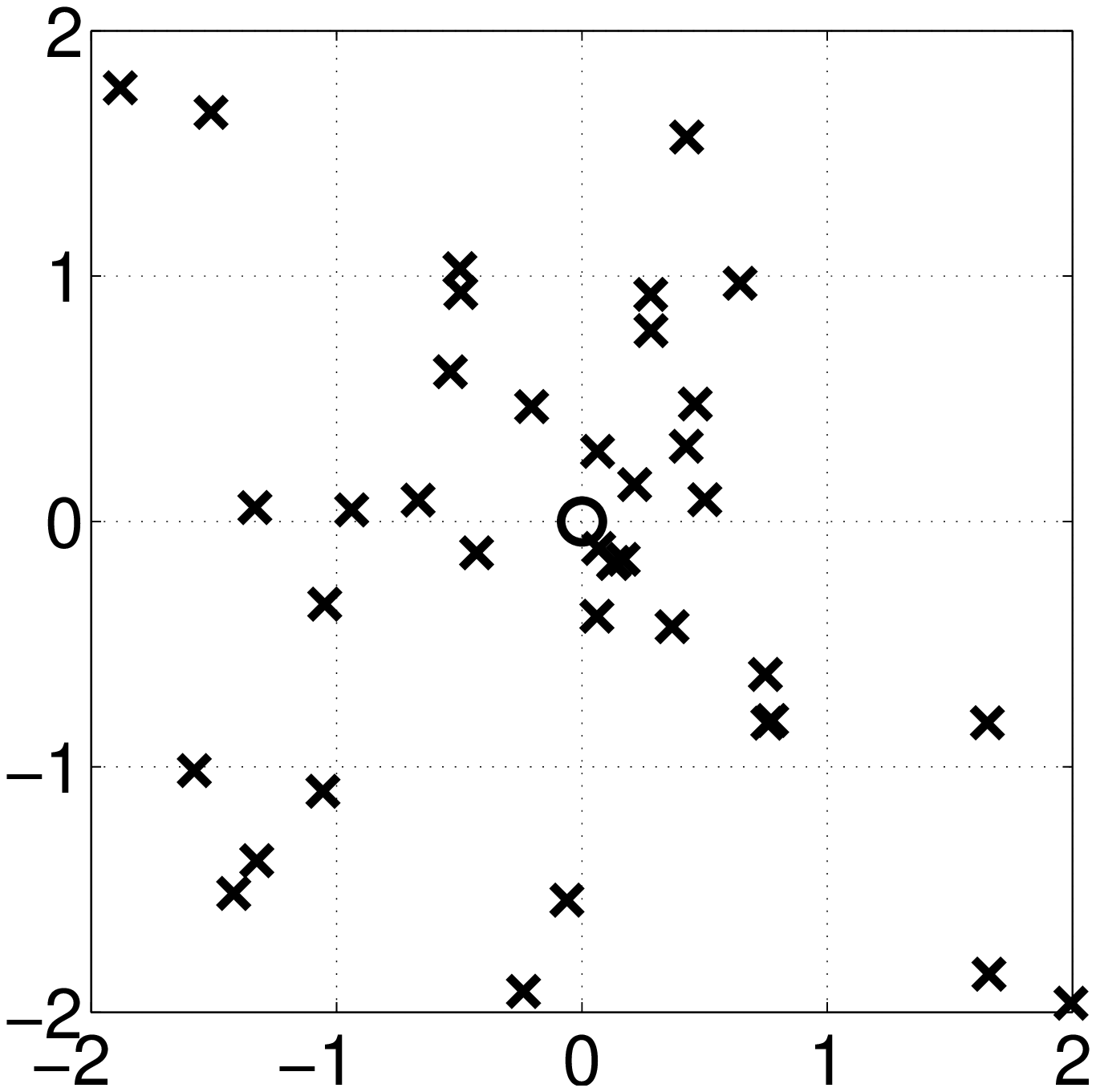}
		}
		\subcaption{$b=-1$\label{subfig:PPPcluster}}
	\end{minipage}
		\begin{minipage}[b]{.28\linewidth}
		\centering
		{
		\includegraphics[width=\linewidth,height=\linewidth]{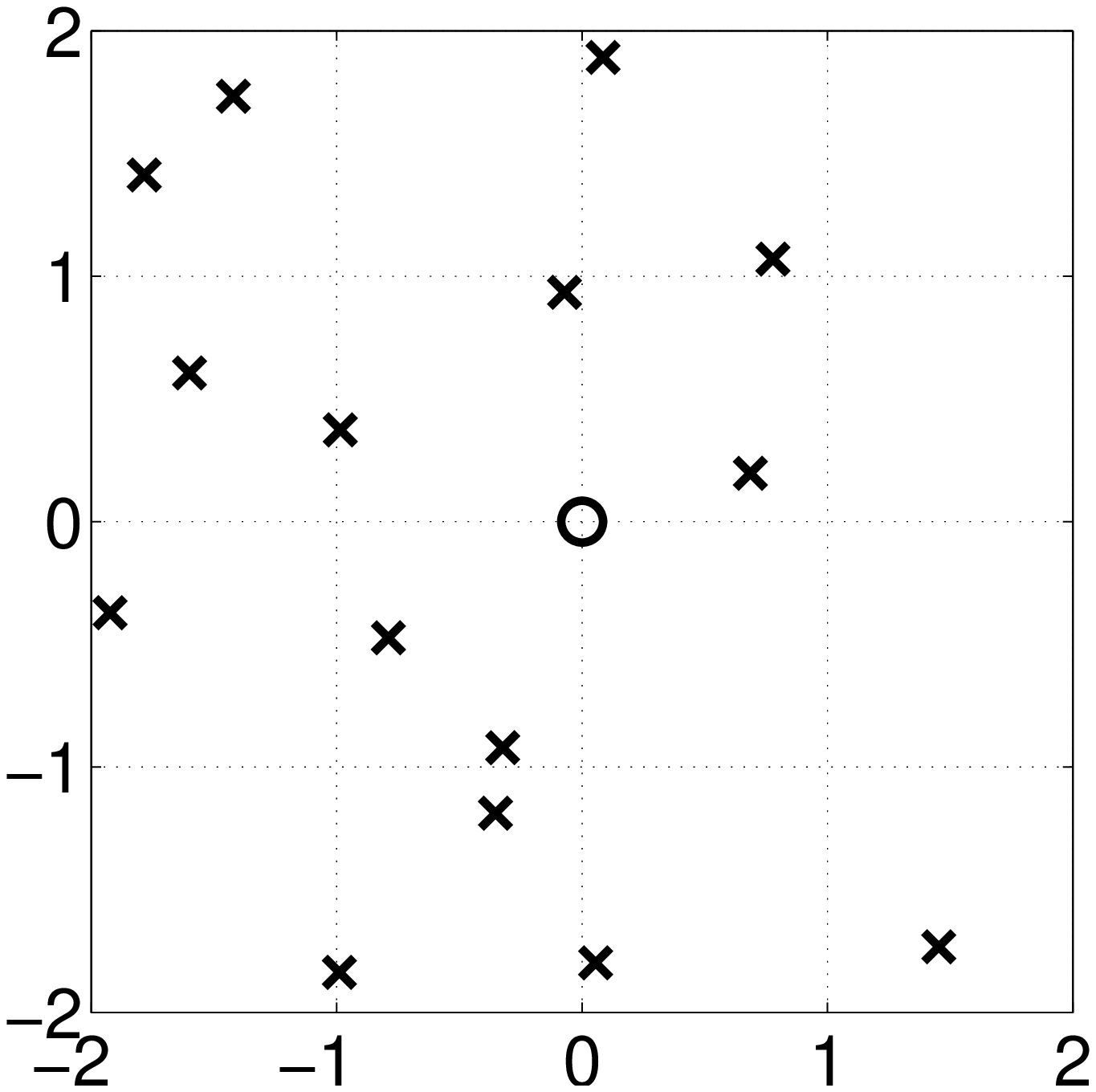}
		}
		\subcaption{$b=0$\label{subfig:PPPuniform}}
	\end{minipage}
	\begin{minipage}[b]{.28\linewidth}
		\centering
		{
		\includegraphics[width=\linewidth,height=\linewidth]{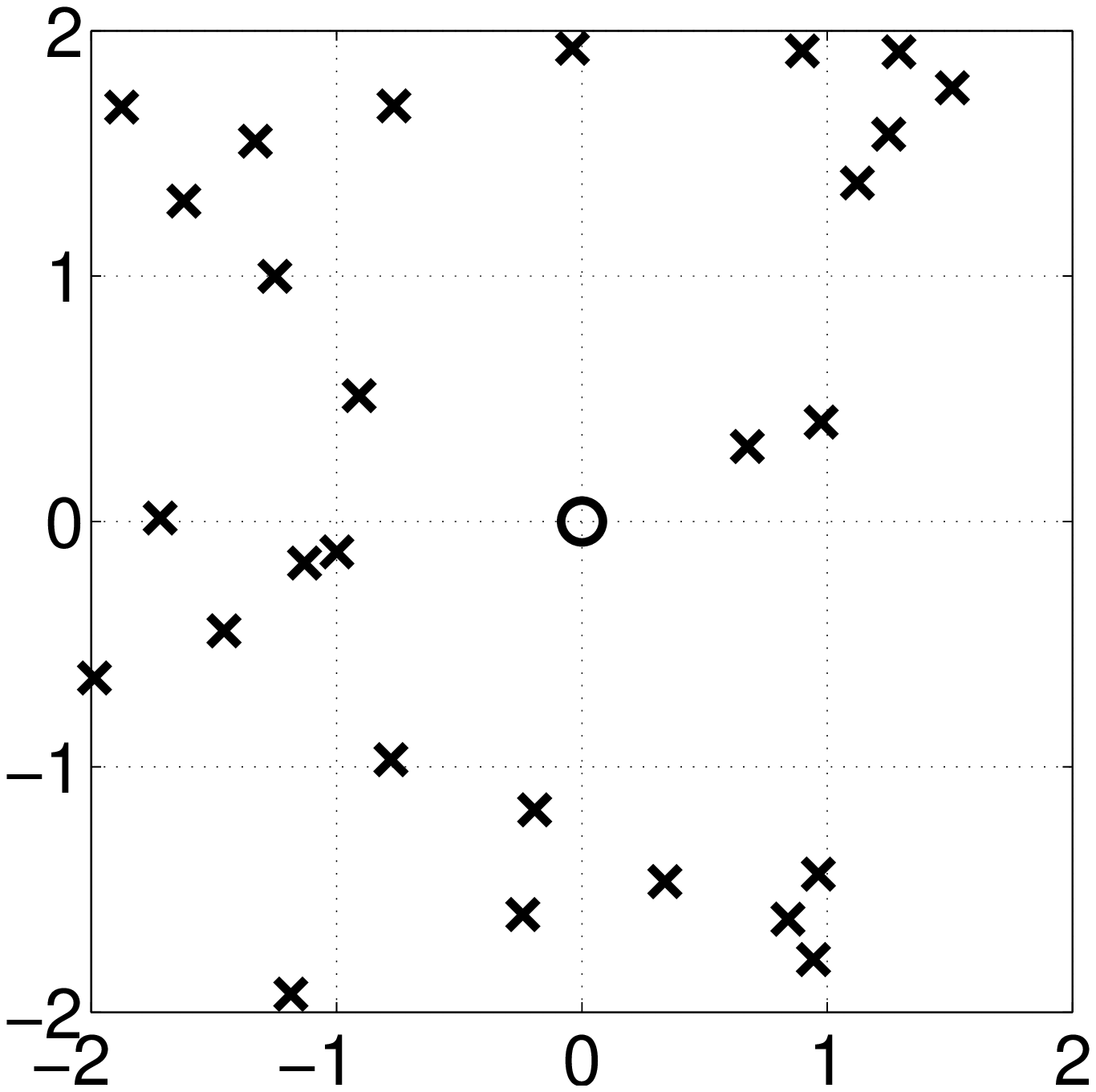}
		}
		\subcaption{$b=1$\label{subfig:PPPinvcluster}}
	\end{minipage}
\caption{Realizations of two non-uniform PPP with intensity function $\lambda(x)=3\|x\|^b$ with different $b$, where $\mathsf{x}$ denotes an active transmitter and $\mathsf{o}$ denotes the receiver at the origin. \label{fig:PPP}}
}
\end{figure}

In this work, we consider a slightly generalized verison of the Poisson
network defined as follows:

\begin{definition}%
The Power-law Poisson Network with Fading (PPNF) is a Poisson network (together with the fading marks)
with density function $\lambda(x)=a\|x\|^b,\; a>0,\;b\in(-d,\alpha-d)$,
where $\|x\|$ is the distance from $x\in\R^d$ to the origin
and $\alpha$ is the path loss exponent.
\label{def:PPNF}
\end{definition}
In Def.~\ref{def:PPNF},
the condition $b\in(-d,\alpha-d)$ is necessary in order
to maintain a finite total received power at $o$ and will be revisited later.
By the definition, we see that
when $b=0$, the PPNF becomes a homogeneous Poisson network with intensity $a$.
Further, the construction of the PPNF provides the flexibility in studying
networks with different clustering properties.
For example, Fig.~\ref{fig:PPP} shows realizations of three 2-d PPNFs with different $b$;
Fig.~\ref{subfig:PPPcluster} represents a network clustered around $o$
whereas the network in Fig.~\ref{subfig:PPPinvcluster} is 
sparse around the receiver at $o$.
In general, the smaller $b$, the more clustered the network is at the origin
with $b=0$ representing the uniform network (\eg Fig.~\ref{subfig:PPPuniform}).

\subsection{SIC Model and Metrics}

Considering the case where all the nodes (users) transmit with unit power,
we recall the following standard signal-to-interference ratio (SIR)-based single
user decoding condition.

\begin{definition}[Standard SIR-based Single User Decoding Condition]
In an interference-limited network,
a particular user at $x\in\Phi$ can be successfully decoded (without SIC) iff
\begin{equation*}
	\textnormal{SIR}_x= \frac{h_x \|x\|^{-\alpha} }{\sum_{y\in\Phi\backslash\{x\}} h_y\|y\|^{-\alpha}}>\theta,
\end{equation*}
where $h_x \|x\|^{-\alpha}$ is the received signal power from $x$, 
$\sum_{y\in\Phi\backslash\{x\}} h_y\|y\|^{-\alpha}$ is the aggregate interference from the other active transmitters,
and $\theta$ is the SIR decoding threshold\footnote{This model will be generalized in Section~\ref{sec:noise} to include noise.}.
\end{definition}

Similarly, in the case of perfect interference cancellation, 
once a user is successfully decoded,
its signal component can be completely subtracted from the received signal.
Assuming the decoding order is always from the stronger users
to the weaker users\footnote{It is straightforward to show that this stronger-to-weaker decoding order maximizes
the number of decodable users and thus the aggregate throughput (defined later) despite the fact that it
is not necessarily the \emph{only} optimal decoding order.},
we obtain the following decoding condition for the case with SIC.

\begin{definition}[SrIR-based Decoding Condition with SIC]
With SIC,
a user $x$ can be decoded if all the users in 
$\mathcal{I}_c=\{y\in\Phi: h_y \|y\|^{-\alpha}>h_x \|x\|^{-\alpha}\}$
are successfully decoded and the signal-to-residual-interference ratio (SrIR)
at $x$ 
\begin{equation*}
\textnormal{SrIR}_x =
\frac{h_x \|x\|^{-\alpha}}{\sum_{y\in\Phi\backslash\{x\}\backslash \mathcal{I}_c} h_y\|y\|^{-\alpha} }> \theta.
\end{equation*}
\end{definition}

Consequently, consider the ordering of all nodes in $\Phi$ such that $h_{x_i}\|x_i\|^{-\alpha}>h_{x_j}\|x_j\|^{-\alpha}$
, $\forall i<j$.\footnote{This ordering is based on received power, which is different from the spatial ordering (based only
on $\Phi$). This is one of the differentiating features of this work compared with the guard-zone-based
analyses in \eg \cite{net:Weber07tit}.}
The number of users that can be successively decoded is $N$ iff
$	h_{x_i}\|x_i\|^{-\alpha}>\theta\sum_{j=i+1}^{\infty} 
	h_{x_j}\|x_j\|^{-\alpha}$, $\forall j\leq N$ and
$h_{x_{N+1}}\|x_{N+1}\|^{-\alpha}\leq\theta\sum_{j=N+2}^{\infty}h_{x_j}\|x_j\|^{-\alpha}$.
Note that the received power ordering is only introduced for analysis purposes.
As is unnecessary, we do not assume that the received power ordering is known \emph{a priori} at the receiver.

One of the goals of this paper is to evaluate $\E [N]$, \emph{i.e.,} the mean number
of users that can be successively decoded, with respect to different system parameters,
and the distribution of $N$ in the form
\begin{equation*}
	p_k\triangleq \P(N\geq k),
\end{equation*}
\emph{i.e.,} the probability of successively decoding at least $k$ users at the origin.
To make the dependence on the point process explicit, we sometimes use $p_k(\hat\Phi)$.

Since SIC is inherently a multiple packet reception (MPR) scheme \cite{ZanellaZorzi12},
we can further define the aggregate throughput (or, sum rate) to be the total information
rate received at the receiver $o$.
Since all the users in the system transmit at the same rate $\log(1+\theta)$,
the sum rate is
\begin{equation}
	R=\E[\log(1+\theta)N]=\log(1+\theta)\E[N].
\label{equ:aggInfoRate}
\end{equation}
Another important goal of this paper is to evaluate $R$ as a function of different system parameters.
Note that this definition of the aggregate throughput counts the information received from all the active
transmitters in the network. Alternatively, one could define an information metric on a subset of
(interested) transmitters and the analyses will be analogous.
One of such instances is the heterogeneous network application discussed in Section~\ref{sec:HetNet}.

\section{The Path Loss Process with Fading (PLPF)\label{sec:PLPF}}

We use the unified framework introduced in \cite{net:Haenggi08tit} to jointly address the randomness from fading and
the random node locations.
We define the path loss process with fading (PLPF) as 
$\Xi\triangleq \{\xi_i=\frac{\|x_i\|^{\alpha}}{h_{x_i}},x_i \in \Phi\}$, where the index $i$ is introduced in the way such that $\xi_i<\xi_j$
for all $i<j$.
Then, we have the following lemma, which
follows from the mapping theorem \cite[Thm. 2.34]{net:mh12}.

\begin{lemma}
The PLPF $\Xi=\{\frac{\|x_i\|^{\alpha}}{h_{x_i}}\}$, where $\{(x_i,h_{x_i})\}$ is a PPNF,
is a one-dimensional PPP on $\R^+$ with intensity measure 
$\Lambda([0,r])=a \delta c_d r^{\beta} \E[h^{\beta}]/\beta$,
where $\delta\triangleq d/\alpha$, 
$\beta\triangleq \delta+b/\alpha\in(0,1)$ and $h$ is a fading coefficient.
\label{lem:mapping}
\end{lemma}

In Lemma~\ref{lem:mapping}, the condition
$\beta\in(0,1)$ corresponds to the condition $b\in(-d,\alpha-d)$
in the definition of the PPNF; it is necessary since
otherwise the aggregate received power at $o$ is infinite almost surely.
More specifically, when $b>\alpha -d$ the intensity measure of the transmitter process
grows faster than the path loss with respect to the network size,
which results in infinite received power at origin,
(\emph{i.e.,} far users contribute infinite power);
when $b<-d$, 
the PLPF is not locally finite (with singularity at $o$),
and thus the number of transmitters that contribute to the received power \emph{more} than any arbitrary value is infinite almost surely,
(\emph{i.e.,} near users contribute infinite power).

Since for all $\xi_i\in\Xi\subset \R^+$, $\xi_i^{-1}$ can be considered as
the $i$-th strongest received power component (at $o$) from the users in $\Phi$,
when studying the effect of SIC, it suffices to just consider the PLPF $\Xi$.
For a PLPF $\Xi$ mapped from $\hat{\Phi}$,
if we let $p_k(\Xi)$ be the probability of successively decoding at least $k$
users in the network $\hat{\Phi}$,
we have the following proposition.

\begin{proposition}[Scale-invariance]
If $\Xi$ and $\bar{\Xi}$ are two PLPFs with intensity measures
$\Lambda([0,r])=r^\beta$ and $\mu([0,r])=C r^\beta$,
respectively, where $C$ is any positive constant,
then
$
		p_k(\Xi)=p_k(\bar{\Xi}),\;\forall k\in\mathbb{N}.
$\label{prop:scale-invariant}
\end{proposition}
\begin{IEEEproof}
Consider the mapping $f(x)=C^{-1/\beta}x$. Then $f(\Xi)$ is a PPP on $\R^+$ with intensity measure $C x^\beta$ of the set $[0,x]$.
Let $\mathcal{N}$ be the sample space of $\Xi$, \emph{i.e.,}
the family of all countable subsets of $\R^+$.
Then, we can define a sequence of indicator functions $\chi_k: \mathcal{N}\to \{0,1\} ,\; k\in\mathbb{N}$,
such that
\begin{equation}
\label{equ:chi_k}
	\chi_k(\phi)=
	\left\{
		\begin{array}{ll}
		1,& \textnormal{if  }	\xi_i^{-1}>\theta I_i,~\forall i\leq k		
		\\
		0,&	\textnormal{otherwise,}
		\end{array}
	\right.
\end{equation}
where $I_i=\sum_{j=i+1}^{\infty}\xi_j^{-1}$, $\phi=\{\xi_i\}$ and $\xi_i<\xi_j,\;\forall i<j$.
Note that $\chi_k(\cdot)$ is scale-invariant, \emph{i.e.,} $\chi_k(\{\xi_i\})=\chi_k(\{C'\xi_i\}),\;\forall C'>0$.
Then, we have
\begin{equation*}
	p_k(\Xi)=
	\mathsf{P}_{\Xi}(Y_k)=
	\E [\chi_k(\Xi)]\peq{a}
	\E [\chi_k (f(\Xi))] \peq{b}
	\E [\chi_k (\bar{\Xi})]=
	\mathsf{P}_{\bar\Xi}(Y_k)=
	p_k(\bar{\Xi}),
\end{equation*}
where $Y_k=\{\phi\in\mathcal{N}:\xi_i^{-1}>\theta I_i,~\forall i\leq k\}$,
$\mathsf{P}_\Xi$ is the probability measure on $\mathcal{N}$ with respect to the distribution of $\Xi$,
(a) is due to the scale-invariance property of $\chi_k(\cdot)$ and (b) is because both $f(\Xi)$
and $\bar{\Xi}$ are PPPs on $\R^+$ with intensity measure $\mu([0,r])=C r^\beta$.
\end{IEEEproof}

Prop.~\ref{prop:scale-invariant} shows that the absolute value of the density is not relevant as long as we restrict our analysis to the power-law density case.
Combining it with Lemma~\ref{lem:mapping},
where it is shown that,
in terms of the PLPF, the only difference introduced by
different fading distributions is a constant factor in the density function,
we immediately obtain the following corollary.

\begin{corollary}[Fading-invariance]
In an interference-limited PPNF,
the probability of successively decoding $k$ users (at the origin) does not depend on the fading distribution
as long as $\E[h^\beta]<\infty$.
\label{cor:fadingdoesntmatter}
\end{corollary}

Furthermore, it is convenient to define a standard PLPF as follows:

\begin{definition}
A \emph{standard PLPF (SPLPF)} $\Xi_\beta$ is a one-dimensional
PPP on $\R^+$ with intensity measure $\Lambda([0,r])=r^\beta$, where $\beta\in (0,1)$.
\label{def:SPLPF}
\end{definition}

Trivally based on Prop.~\ref{prop:scale-invariant}
and Cor.~\ref{cor:fadingdoesntmatter}, the following fact significantly
simplifies the analyses in the rest of the paper.

\begin{fact}
The statistics of $N$ in a PPNF are identical to those of $N$ in $\Xi_\beta$ for
any fading distribution and any values of $a$, $b$, $d$, $\alpha$,
with $\beta=\delta+b/\alpha=(d+b)/\alpha$.
\label{fact:SPLPF}
\end{fact}

\section{Bounds on the Probability of Successive Decoding\label{sec:boundsonpk}}

Despite the unified framework introduced in Section~\ref{sec:PLPF},
analytically evaluating $p_k$
requires the joint distribution of the received powers from the $k$
strongest users and the aggregate interference from the rest of the network,
which is daunting even for 
the simplest case of a one-dimensional homogeneous PPP.
In this section, we derive bounds on $p_k$.
Due to the technical difficulty of deriving a bound that is tight for all network parameters,
we provide different tractable bounds tight for different system parameters.
These bounds complement each other and collectively provide insights on how $p_k$ depends on different system parameters.
The relations between different bounds are summarized in Table~\ref{tab:addlabel} at the end of this section.

\subsection{Basic Bounds}

The following lemma introduces basic upper and lower bounds on $p_k$
in terms of the probability of decoding the $k$-th strongest user assuming the $k-1$
strongest users do not exist.
Although not being bounds in closed-form,
the bounds %
form the basis for the bounds introduced later.

\begin{lemma}
In a PPNF, the probability of successively decoding $k$ users is bounded as follows:
\begin{itemize}
\item
$p_k \geq
(1+\theta)^{-\frac{\beta k(k-1)}{2}}\P(\xi_k^{-1}>\theta I_{k})$
\item
$p_k \leq \theta^{-\frac{\beta k(k-1)}{2}}\P(\xi_k^{-1}>\theta I_{k})$
\end{itemize}
where $\Xi_\beta=\{\xi_i\}$ is the corresponding SPLPF and $I_{k} \triangleq \sum_{j=k+1}^\infty \xi_j^{-1}$.
\label{lem:d-dim}
\end{lemma}

\begin{IEEEproof}
See App.~\ref{app:d-dim}.
\end{IEEEproof}

The idea behind of Lemma~\ref{lem:d-dim} is to first decompose $p_k$ by Bayes' rule
into $\P(\xi_i>I_i,\; \forall i\in[k-1] \mid \xi_k^{-1}>I_k)\P(\xi_k^{-1}>I_k)$,
and then to bound the first term.
An important observation is that conditioned on $\xi_k$, the distribution
of $\xi_i/\xi_k,\; \forall  i<k$ is the same as that of the $i$-th order statistics of $k-1$ iid
random variable with cdf $F(x)=x^\beta\mathsf{1}_{[0,1]}(x)$.
This observation allows us to bound $\P(\xi_i>I_i,\; \forall i\in[k-1] \mid \xi_k^{-1}>I_k)$
using tools from the order statistics
of uniform random variables \cite{bk:OrderStatistics} since $F(x)$ is also the cdf of $U^{\frac{1}{\beta}}$,
where $U$ is a uniform random variable with support $[0,1]$.

Since $\lim_{\theta\to\infty} \frac{\theta}{1+\theta} = 1$,
it is observed that both the upper and lower bounds in Lemma~\ref{lem:d-dim}
are asymptotically tight when $\theta\to\infty$, for all $\beta\in(0,1)$ and $k\in\mathbb{N}$.
Further, as will be shown later, the bounds are quite informative for
moderate and realistic values of $\theta$.

The importance of Lemma~\ref{lem:d-dim}
can be illustrated by the following attempt of expressing
$p_k$ in a \emph{brute-force} way.
Letting $f_{\xi_1,\xi_2,\cdots,\xi_k, I_k}(\cdot)$ be the joint distribution (pdf)
of $\xi_1, \xi_2, \cdot, \xi_k$ and $I_k$, we have
\begin{equation}
	p_k = \int_0^{\infty}\int\limits_0^{\frac{1}{\theta y}}\int\limits_0^{\frac{1}{\theta (y+x_k^{-1})}}
	\int\limits_0^{\frac{1}{\theta (y+\sum_{i=k-1}^k x_i^{-1})}} \cdots
	\int\limits_0^{\frac{1}{\theta (y+\sum_{i=2}^k x_i^{-1})}}
	f_{\xi_1,\xi_2,\cdots,\xi_k, I_k}(x_1, x_2, \cdots, x_k, y)\d x_1 \d x_2 \cdots \d x_k \d y.
\label{equ:brutal-force}
\end{equation}
There are two main problems with using \eqref{equ:brutal-force} to study the performance of SIC:
First, the joint distribution of $f_{\xi_1,\xi_2,\cdots,\xi_k, I_k}(\cdot)$
is hard to get as pointed out also in \cite{net:Weber07tit}.
Second, even if the joint distribution is obtained by possible numerical inverse-Laplace
transform, the $k+1$ fold integral is very hard to be numerically calculated,
and it is very likely that the integration is even more numerically intractable than
a Monte Carlo simulation\footnote{In this case, it is desirable to integrate by Monte Carlo methods.
But that can only bring down the complexity to the level of simulations.}.
Even if the above two problems are solved,
the closely-coupled $k+1$ fold integral in \eqref{equ:brutal-force} is
very hard to interpret, and thus offers little design insights on the performance of SIC.

\subsection{The Lower Bounds\label{subsec:TheLowerBounds}}

\subsubsection{High-rate lower bound}
Lemma~\ref{lem:d-dim} provides bounds on $p_k$ as a function of $\P(\xi_k^{-1}>\theta I_k)$.
In the following, we give the high-rate lower bounds\footnote{The high-rate lower bound also holds in the low-rate case, \emph{i.e.,} $\theta$ is small. The bound is named as such since in the low-rate case we will provide another (tighter) bound.} by lower bounding $\P(\xi_k^{-1}>\theta I_k)$.

\begin{lemma}\label{lem:kNN}
The $k$-th smallest element in $\Xi_\beta$, $\xi_k$, has pdf
\begin{equation*}
	f_{\xi_k}(x)=\frac{ \beta x^{k\beta-1}}{\Gamma(k)}\exp(- x^\beta).
\end{equation*}
\end{lemma}

Thanks to the Poisson nature of $\Xi$ (Lemma~\ref{lem:mapping}),
the proof of Lemma~\ref{lem:kNN} is analogous to the one of \cite[Thm. 1]{net:Haenggi05tit}
where the result is only about the distance (fading is not considered).

\begin{lemma}
For $\Xi_\beta=\{\xi_i\}$,
$\P(\xi_k^{-1}>\theta I_k )$ is lower bounded by
\begin{equation*}
	\Delta_1(k) \triangleq
	 \frac{1}{\Gamma(k)}\left(	\gamma\left(k,\frac{1-\beta}{\theta\beta}\right)
	-\frac{\theta\beta}{1-\beta}\gamma\left(k+1,\frac{1-\beta}{\theta\beta}\right)	\right),
\end{equation*}
where $\gamma(\cdot,\cdot)$ is the \emph{lower} incomplete gamma function.\label{lem:boundconditionalpk}
\end{lemma}

The proof the Lemma~\ref{lem:boundconditionalpk} is a simple application of the Markov inequality
and can be found in App.~\ref{app:proofsofLHbounds}.
In principle, one could use methods similar to the one in the proof of Lemma~\ref{lem:boundconditionalpk}
to find the higher-order moments of $I_\rho$ and then obtain tighter bounds by applying
inequalities involving these moments, \emph{e.g.,} the Chebyshev inequality.
However, these bounds cannot be expressed in closed-form, and the improvements are marginal.

Combining Lemmas~\ref{lem:d-dim}~and~\ref{lem:boundconditionalpk},
we immediately obtain the following proposition.

\begin{proposition}[High-rate lower bound]\label{prop:pk_hr_lb}
In the PPNF, $p_k \geq
(1+\theta)^{-\frac{\beta k(k-1)}{2}}\Delta_1(k)$.
\end{proposition}

Since $\Delta_1(k)$ is monotonically decreasing with $k$,
the lower bound in Prop.~\ref{prop:pk_hr_lb} decays super-exponentially with $k^2$.

\subsubsection{Low-rate lower bound}
The lower bound in Prop.~\ref{prop:pk_hr_lb} is tight for large $\theta$.
However, it becomes loose when $\theta$ is small.
This is because Prop.~\ref{prop:pk_hr_lb} estimates $p_k$ by approximating
the relation between $\xi_i$ and $I_i$ with the relation between
$\xi_i$ and $\xi_{i+1}$.
This approximation is accurate when $\xi_{i+1}^{-1} \approx \theta I_{i+1}$.
But when $\theta \rightarrow 0$, $\xi_{i+1}^{-1} \gg \theta I_{i+1}$
happens frequently, making the bound loose.
The following proposition provides an alternative lower bound particularly tailored for the small $\theta$ regime.

\begin{proposition}[Low-rate lower bound]
\label{prop:pk_lr_lb}
In the PPNF, 
for $k<1/\theta +1$, $p_k$ is lower bounded 
by
\begin{equation*}
 \underline{p}^{\textnormal{LR}}_k
 \triangleq 	 \frac{1}{\Gamma(k)}\left(	\gamma\left(k,\frac{1-\beta}{\tilde{\theta}\beta}\right)
	-\frac{\tilde{\theta}\beta}{1-\beta}\gamma\left(k+1,\frac{1-\beta}{\tilde{\theta}\beta}\right)	\right),
\end{equation*}
where ${\textnormal{LR}}$ means \emph{low-rate} and
$\tilde{\theta}\triangleq \frac{\theta}{1-(k-1)\theta}$.
\end{proposition}

In order to avoid the limitation of estimating $I_i$ by $\xi_i$ when $\theta\to 0$,
the low-rate lower bound in Prop.~\ref{prop:pk_lr_lb} is not based on Lemma~\ref{lem:d-dim}.
Instead, we observe that $I_i=\sum_{j=i+1}^k \xi_{j}^{-1} + I_k < (k-i) \xi_i^{-1}+I_k, \;\forall i<k$
and thus the probability of $\xi_i^{-1}>\theta I_i$ can be estimated by the joint distribution of $\xi_i$ and $I_k$.
Recursively apply this estimate for all $i<k$ leads to the bound as stated.
The proof of Prop.~\ref{prop:pk_lr_lb} is given in App.~\ref{app:proofsofLHbounds}.
Note that the bound in Prop.~\ref{prop:pk_lr_lb} is only defined for $k<1/\theta +1$.
Yet, in the low rate regime ($\theta\to 0$), this is not a problem.
As will be shown in Section~\ref{sec:EN},
when $\theta \rightarrow 0$, this bound behaves much better than
the one in Prop.~\ref{prop:pk_hr_lb}.

\subsection{The Upper Bound\label{subsec:TheUpperBound}}

Similar to the high-rate lower bound, 
we derive an upper bound by upper bounding $\P(\xi_k^{-1} >\theta I_k)$.

\begin{lemma}
For $\Xi_\beta=\{\xi_i\}$,
$\P(\xi_k^{-1}>\theta I_k)$ is upper bounded by
\begin{equation*}
\Delta_2(k) \triangleq
 \bar\gamma(k,1/c)+\frac{e}{(1+c)^k}\bar\Gamma(k,1+1/c),
\end{equation*}
where $c=\theta^\beta \gamma(1-\beta,\theta)-1+e^{-\theta}$,
$\bar\gamma(z,x)=\frac{\gamma(z,x)}{\Gamma(z)}$
and $\bar\Gamma(z,x)=\frac{\Gamma(z,x)}{\Gamma(z)}$ are the \emph{normalized} \emph{lower} and \emph{upper}
incomplete gamma function,
and $\Gamma(\cdot,\cdot)$ is the \emph{upper} incomplete gamma function.\label{lem:upboundconditionalpk2}
\end{lemma}

The proof of Lemma~\ref{lem:upboundconditionalpk2} (see App.~\ref{app:proofsofLHbounds}) relies on the idea of
constructing an artificial Rayleigh fading coefficient and compare the outage probability
in the original (non-fading) case and the fading case. 
Combining Lemmas~\ref{lem:upboundconditionalpk2}~and~\ref{lem:d-dim}
yields the following proposition.

\begin{proposition}[Combined upper bound]\label{prop:pk_c_ub}
In the PPNF,
we have $	p_k\leq \overline p_k \triangleq	\bar{\theta}^{-\frac{\beta}{2}k(k-1)}  \Delta_2(k)$, 
where $\bar{\theta}=\max\{\theta,1\}$.
\end{proposition}

For $\theta>1$, similar to the high-rate lower bound in Prop.~\ref{prop:pk_hr_lb},
the upper bound in Prop.~\ref{prop:pk_c_ub} decays 
super-exponentially with $k^2$, \emph{i.e.,} $-\log \overline p_k \propto k^2$,
which suggests that, in this regime, 
the marginal gain of adding SIC capability 
(\emph{i.e.,} the ability of successively cancelling more users)
diminishes very fast.

\subsection{The Sequential Multi-user Decoding (SMUD) Bounds\label{subsec:AlternativeBounds}}

The bounds derived in Sections~\ref{subsec:TheLowerBounds}~and~\ref{subsec:TheUpperBound}
apply to all $\theta>0$.
This subsection provides an alternative set of bounds constructed based on a different idea.
These bounds are typically much tighter than the previous bounds in the sequential multi-user decoding (SMUD) regime
defined as follows.

\begin{definition}
A receiver with SIC capability is in the sequential multi-user decoding (SMUD) regime if the
decoding threshold $\theta\geq 1$.
\end{definition}

It can be observed that in the SMUD regime multiple packet reception (MPR)
can be only carried out with the help of SIC,
whereas outside this regime, \ie $\theta<1$, MPR is possible without SIC, \ie by parallel decoding
(this argument is made rigorous by Lemma~\ref{lem:k-strongest} in App.~\ref{app:Pxik>thetaIk}).
This important property of the SMUD regime enables us to show the following
(remarkable) result which gives a closed-form expression for 
$\P(\xi_k^{-1}>\theta I_k)$.

\begin{theorem}

For $\theta\geq 1$,
\begin{equation}
	\P(\xi_k^{-1}>\theta I_k)=
	\frac{1}{\theta^{k\beta}\Gamma(1+k\beta)\big(\Gamma(1-\beta)\big)^{k}},
\label{equ:generalPxik}
\end{equation}
where $\Gamma(\cdot)$ is the gamma function.
Moreover, the RHS of (\ref{equ:generalPxik}) is an upper bound on $\P(\xi_k^{-1}>\theta I_k)$
when $\theta < 1$.
\label{thm:Pxik>thetaIk}
\end{theorem}

With details of the proof in App.~\ref{app:Pxik>thetaIk},
the main idea of Thm.~\ref{thm:Pxik>thetaIk} lies in the observation that, in the SMUD
regime, there can be \emph{at most} one $k$-element user set,
where the received power from any one of the $k$ users is larger
than $\theta$ times the interference from the rest of the network.
This observation, combined with the fading-invariance property shown in Cor.~\ref{cor:fadingdoesntmatter},
enables us to separate the $k$ intended users from the rest of the network
under \emph{induced} (artificial) fading without worrying about overcounting.
Conversely, with $\theta<1$, overcounting cannot be prevented, which
is why the same method results in an upper bound.

\begin{figure}[t]
\centering
\psfrag{aaaaaaaaaaaaa}{$\P(\xi_k^{-1}>\theta I_k)$}
\psfrag{a}{$k=1$}
\psfrag{b}{$k=5$}
\includegraphics[width=0.5\linewidth]{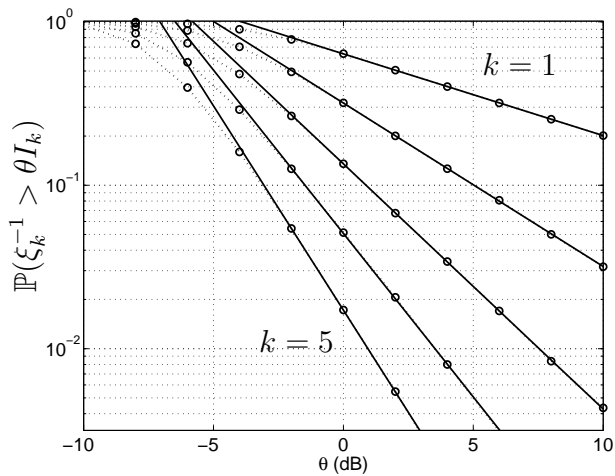}
\caption{Comparison of $\P(\xi_k^{-1}>\theta I_k)$ between simulation and the analytical value according to Cor.~\ref{cor:beta=.5} for $k=1,2,3,4,5$.}
\label{fig:pxik_betapt5}
\end{figure}

Combining Thm.~\ref{thm:Pxik>thetaIk} with Lemma~\ref{lem:d-dim},
we obtain another set of bounds on $p_k$.

\begin{proposition}[SMUD bounds]\label{prop:altbounds_pk}
For $\theta\geq 1$ and $\Xi_\beta=\{\xi_i\}$, we have
\begin{equation*}
 p_k \geq   
	\frac{1}{(1+\theta)^{\frac{\beta}{2} k (k-1)}\theta^{k\beta}\Gamma(1+k\beta)\big(\Gamma(1-\beta)\big)^{k}}
\label{equ:nbpklb}
\end{equation*}
and
\begin{equation*}
	p_k
	\leq 
	\frac{1}{\theta^{\frac{\beta}{2} k(k+1)} \Gamma(1+k\beta)\big(\Gamma(1-\beta)\big)^{k}}.
\label{equ:nbpkub_theta>=1}
\end{equation*}
More generally, for all $\theta>0$, we have
\begin{equation}
	p_k
	\leq 
	\frac{1}{\bar\theta^{\frac{\beta}{2} k(k-1)} \theta^{k\beta}\Gamma(1+k\beta)\big(\Gamma(1-\beta)\big)^{k}},
\label{equ:nbpkub}
\end{equation}
where $\bar\theta=\max\{\theta,1\}$.
\end{proposition}

Note that the SMUD upper bound in Prop.~\ref{prop:altbounds_pk} is valid also for $\theta<1$.
The name of the bounds only suggests that these bounds are tightest in the SMUD regime.

\subsection{Two General Outage Results}

Taking $k=1$, we obtain the following corollary of Thm.~\ref{thm:Pxik>thetaIk},
which gives the exact probability of decoding the strongest user in a PPNF for $\theta>1$ and a general upper bound of the probability of decoding the strongest user.

\begin{corollary}
For $\theta\geq 1$, we have
\begin{equation}
	p_1=\P(\xi^{-1}_1>\theta I_1) = \frac{\sinc \beta}{\theta^\beta},
\label{equ:p1}
\end{equation}
and the RHS is an upper bound on
$\P(\xi_1^{-1}>\theta I_1)$
when $\theta < 1$.
\label{cor:coverageProb_highSIR}
\end{corollary}

It is worth noting that the closed-form expression in Cor.~\ref{cor:coverageProb_highSIR}
has been discovered in several special cases.
For example, \cite{net:Dhillon12jsac} derived the equality part of (\ref{equ:p1})
in the Rayleigh fading case,
and \cite{PrasannaMadhusudhanan2012} showed that the equality is true for arbitrary
fading distribution.
However, none of the existing works derives the results in Cor.~\ref{cor:coverageProb_highSIR}
in as much generality as here.
More precisely, we proved that (\ref{equ:p1}) holds for arbitrary fading (including the non-fading case)
in $d$-dimensional PPNF (including non-uniform user distribution).

When $\beta=\frac{1}{2}$, (\ref{equ:generalPxik})
can be further simplified, and we have the following corollary.

\begin{corollary}
When $\beta = 1/2$,
\begin{equation}
	\P(\xi_k^{-1}>\theta I_k) =  \frac{1}{(\pi\theta)^{\frac{k}{2}}\Gamma(\frac{k}{2}+1)},
\label{equ:beta=.5}
\end{equation}
and the RHS is an upper bound on $\P(\xi_k^{-1}>\theta I_k)$
when $\theta<1$.
\label{cor:beta=.5}
\end{corollary}

Fig.~\ref{fig:pxik_betapt5} compares the \eqref{equ:beta=.5} with
simulation results for $k=1,2,3,4,5$.
We found that the estimate in Cor.~\ref{cor:beta=.5} is 
quite accurate for $\theta>-4\textnormal{\;dB}$,
which is consistent with the observation in \cite{net:Dhillon12jsac},
where only the case $k=1$ is studied.

\subsection{Comparison of the Bounds}

\begin{figure}[t]
\centering
\psfrag{a}{$k=1$}
\psfrag{b}{$k=2$}
\psfrag{c}{\hspace{-5pt}$k=3$}
\psfrag{p}{$p_k$}
\psfrag{k}{}
\includegraphics[width=0.5\linewidth]{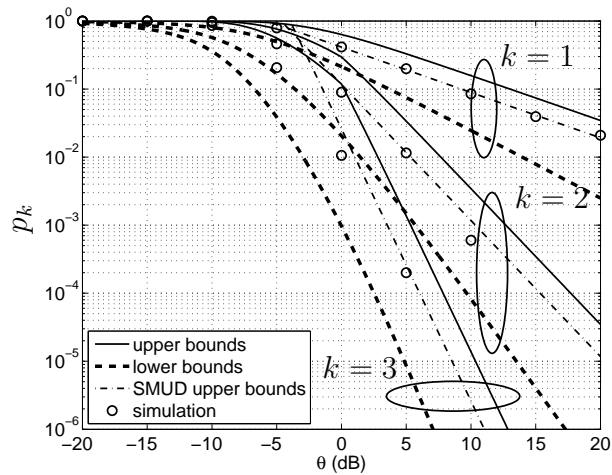}
\caption{Combined upper bound (Prop.~\ref{prop:pk_c_ub}), high-rate lower bound (Prop.~\ref{prop:pk_hr_lb}), and SMUD upper bound (Prop.~\ref{prop:altbounds_pk}) for $p_k$ ($k=1,2,3$, from top to bottom) in a 2-d uniform network with path loss exponent $\alpha=3$.}
\label{fig:SICpk123}
\end{figure}

Focusing on $k=1,2,3$, Fig.~\ref{fig:SICpk123}
plots the combined upper bounds, high-rate lower bounds, SMUD upper bounds as a function of $\theta$.
We see that $p_k$ decays very rapidly with $\theta$, especially when $k$ is large,
which suggests that the benefit of decoding many users can
be very small under high-rate codes.

As is shown in the figure,
the SMUD bounds are generally tighter than the combined upper bound.
However, these bounds are less informative when $\theta\ll 1$,
where the upper bound exceeds one at about $-5$\;dB.
After that, we have to rely on the combined upper bound to estimate $p_k$.
Note that the combined upper bound behaves slightly differently
for $\theta>1$ and $\theta<1$ when $k>1$.
This is because the combined upper bound in Prop.~\ref{prop:pk_c_ub} 
becomes $\Delta_2(k)$ when $\theta<1$.
More precisely, the combined upper bound ignores the ordering among
the $k$ strongest users and only considers $\P(\xi_k^{-1}>\theta I_k)$
when $\theta<1$.

\begin{figure}[t]
\centering
{
	\begin{minipage}[t]{.47\linewidth}
		\centering
		{
		\psfrag{pk}{$p_k$}
		\includegraphics[width=\linewidth]{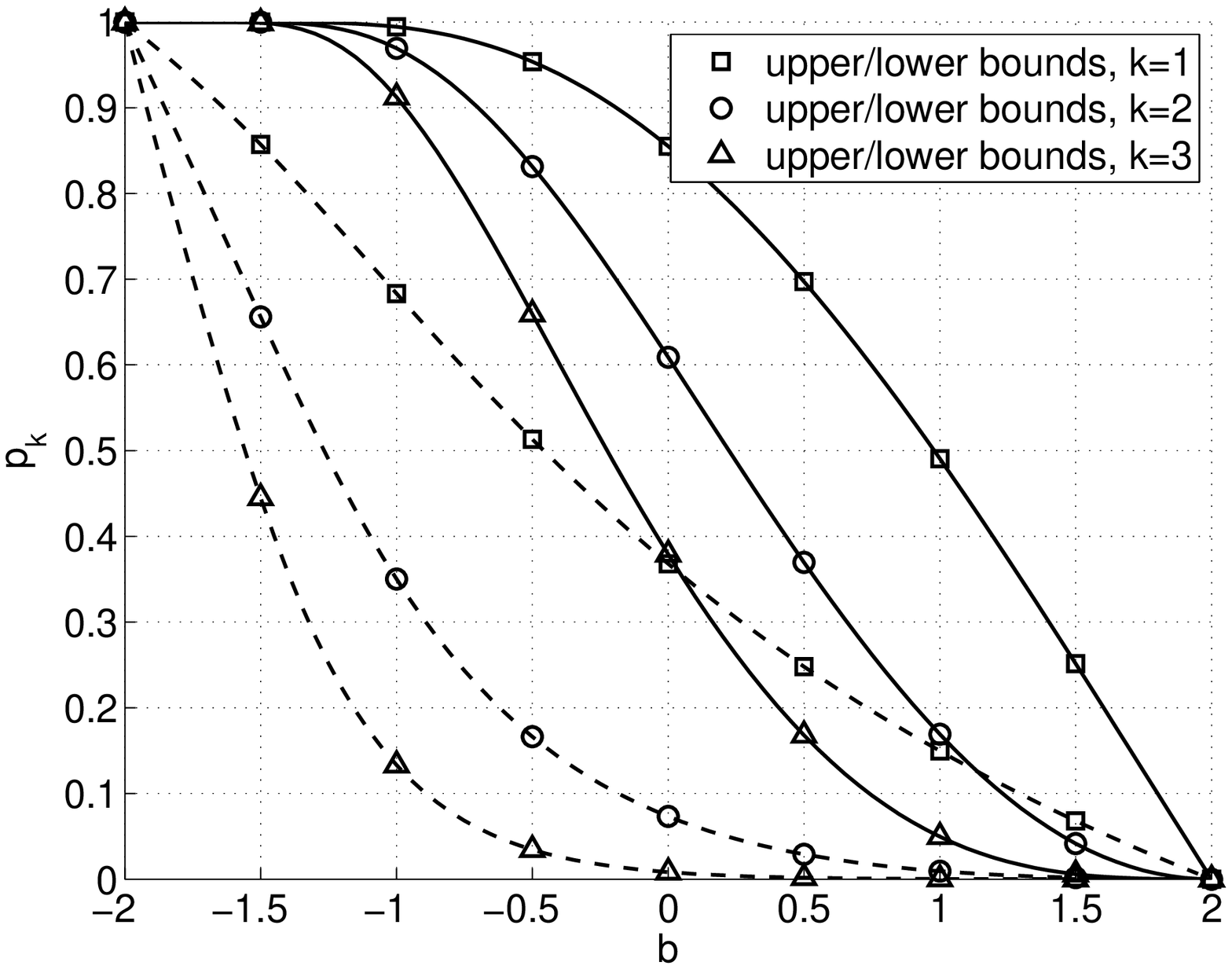}
		\subcaption{Combined upper bound and high-rate
		lower bound.\label{subfig:SICpk123clustering}}
		}		
	\end{minipage}
	\begin{minipage}[t]{.47\linewidth}
		\centering
		{
		\includegraphics[width=\linewidth]{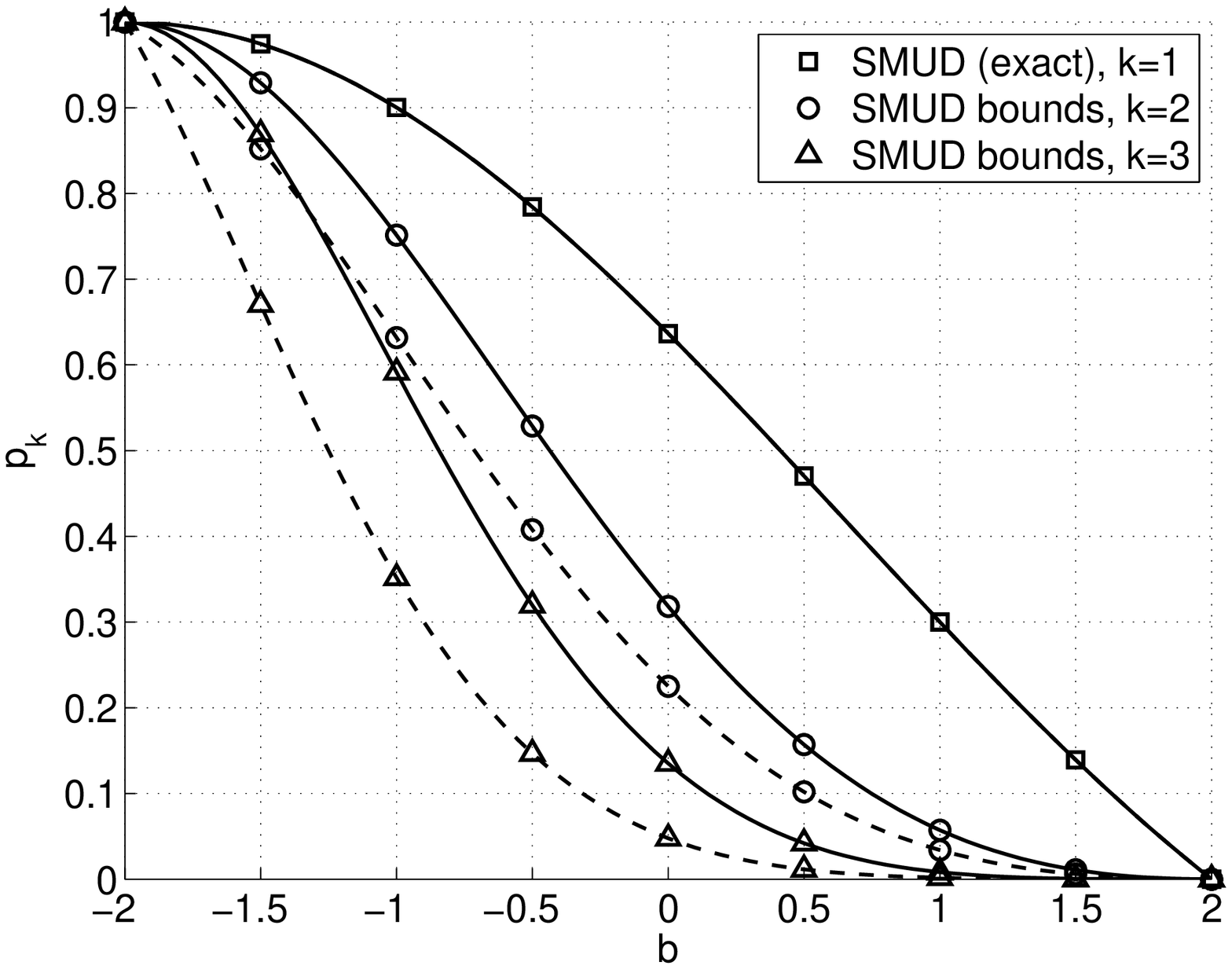}
		\subcaption{SMUD estimates.\label{subfig:SICpk123clustering_alt}}
		}		
	\end{minipage}
	\vspace{-1cm}
\caption{Upper and lower bounds for $p_k$ ($k=1,2,3$) in a 2-d network with
with path loss exponent $\alpha=4$, $\theta=1$ and density function $\lambda(x)=a\|x\|^b$.
$b=0$ is the uniform case. Upper bounds are in solid lines, and
lower bounds in dashed lines. \label{fig:SICpk123clustering}}
}
\end{figure}

The bounds derived above highlight the impact
of clustering on the effectiveness of SIC.
Fig.~\ref{subfig:SICpk123clustering} compares the bounds on probability
of successively decoding 1, 2 and 3 users for different network clustering parameters $b$,
using the combined upper and high-rate lower bounds derived in Sections~\ref{subsec:TheLowerBounds} and \ref{subsec:TheUpperBound}.
The corresponding SMUD bounds on the same quantities derived in Section~\ref{subsec:AlternativeBounds}
are plotted in Fig.~\ref{subfig:SICpk123clustering_alt},
where the upper and lower bounds for the case $k=1$ are both tight and overlapping.
Comparing Figs.~\ref{subfig:SICpk123clustering}~and~\ref{subfig:SICpk123clustering_alt},
we find the SMUD bounds are a huge improvement over the combined upper bound and high-rate
lower bound despite its limitation.
While the bounds in Fig.~\ref{fig:SICpk123clustering}
are derived using rather different techniques and provide different levels of tightness
for different values of $\theta$,
both figures capture the important fact that
{\em the more clustered the network, the more useful SIC.}

Table~\ref{tab:boundcomparison} summarizes and compares the three lower bounds
and two upper bounds derived in this section.
In general, the SMUD bounds are the best estimates if $\theta\geq 1$.
However, there is no SMUD lower bound defined for $\theta< 1$ and
the SMUD upper bound becomes trivial (exceeds one) for $\theta\ll 1$.
This is the reason why we need the other bounds to complement the SMUD bounds.
\begin{table}[htbp]
  \centering
  \caption{Comparison of Different Bounds on $p_k$: LR stands for low-rate; HR stands for high-rate.\label{tab:boundcomparison}}
    \begin{tabular}{|c|c|c|c|c|c|}
    \hline
    \multirow{2}[4]{*}{} & \multicolumn{3}{c|}{Lower Bounds} & \multicolumn{2}{c|}{Upper Bounds} \\
\cline{2-6}          & HR    & LR    & SMUD  & Combined & SMUD \\
    \hline
    Given in & Prop.~\ref{prop:pk_hr_lb} & Prop.~\ref{prop:pk_lr_lb} & Prop.~\ref{prop:altbounds_pk} & Prop.~\ref{prop:pk_c_ub} & Prop.~\ref{prop:altbounds_pk} \\
    \hline
    Based on the Basic bounds & Yes   & No    & Yes   & Yes   & Yes \\
    \hline
    Valid/nontrivial when & $\theta\in\R^+$ & $k<\theta^{-1} +1$ & $\theta\geq 1$ & $\theta\in\R^+$ & $\theta\not\ll 1$ \\
    \hline
    $\lim_{\theta\to 0}(\cdot)=1$ & Yes   & Yes   & N/A   & Yes    & No \\
    \hline
    Typical Best Estimate Region &  $\theta\not\ll 1$ & $\theta\ll 1$ & $\theta\geq 1$ & $\theta\ll 1$ & $\theta\not\ll 1$ \\
    \hline
    \end{tabular}%
  \label{tab:addlabel}%
\end{table}%

\section{The Expected Gain of SIC\label{sec:EN}}

\subsection{The Mean Number of Successively Decoded Users}

With the bounds on $p_k$,
we are able to derive bounds on $\E[N]$, the expected 
number of users that can be successively decoded in the system,
since $\E[N]=\sum_{k=1}^\infty p_k$.

\begin{proposition}
In the PPNF, we have
$\E[N] \geq\sum_{k=1}^{K} (1+\theta)^{-\frac{\beta}{2}k(k-1)} \Delta_1(k)$
for all $K\in\mathbb{N}$. %
\label{prop:ENlb}
\end{proposition}

On the one hand,
Prop.~\ref{prop:ENlb} follows directly from Prop.~\ref{prop:pk_hr_lb}
when $K\rightarrow \infty$.
On the other hand,
since for large $\theta$, $p_k$ decays very fast with $k$,
a tight approximation can be obtained for small integers $K$.
In fact, the error term $\sum_{k=K+1}^{\infty} (1+\theta)^{-\frac{\beta}{2}k(k-1)}\Delta_1(k)$
can be upper bounded as
\begin{align}
\sum_{k=K+1}^{\infty} (1+\theta)^{-\frac{\beta}{2}k(k-1)}\Delta_1(k)	\notag	
&\leq	\Delta_1(K)\sum_{k=K+1}^{\infty} (1+\theta)^{-\frac{\beta}{2}k(k-1)}	\notag	\\
&\leq \Delta_1(K)\int_{K}^{\infty} (1+\theta)^{-\frac{\beta}{2}x(x-1)}\d x	\notag \\
&= \frac{(1+\theta)^{\frac{\beta}{8}}\Delta_1(K)\sqrt{\pi}}{\sqrt{2\beta\log(1+\theta)}}\operatorname{erfc}\left((K-\frac{1}{2})
			\sqrt{\frac{\beta}{2}\log(1+\theta)}\right),	\label{equ:err_bd}
\end{align}
where $\erfc(\cdot)$ is the complementary error function.
By inverting (\ref{equ:err_bd}), one can control the numerical error introduced by choosing an finite $K$.
Due to the tail property of complementary error function 
and the monotonicity of $\Delta_1(k)$,
it is easy to show that
the error term decays super-exponentially with $K^2$ when $K\gg 1$ and thus a finite $K$
is a good approximation for the case $K\rightarrow \infty$ .

On the other hand, 
\begin{equation}
\frac{(1+\theta)^{\frac{\beta}{8}}\Delta_1(K)\sqrt{\pi}}{\sqrt{2\beta\log(1+\theta)}}\operatorname{erfc}\left((K-\frac{1}{2})
			\sqrt{\frac{\beta}{2}\log(1+\theta)}\right)\\
\sim \sqrt{\frac{\pi}{2\beta}} \theta^{-\frac{1}{2}},~\textnormal{as}~\theta\to 0,
\label{equ:err_bd_asymp}
\end{equation}
where we use the fact that
$\lim_{\theta\to 0} \Delta_1(K)=1$ and $\lim_{x\to 0}\operatorname{erfc} (x)=1$.
(\ref{equ:err_bd_asymp}) suggests that when $\theta\rightarrow 0$,
for any finite $K$, the error \emph{may} blow up quickly,
which is verified numerically.
Therefore, in the small $\theta$ regime, we need another, tighter, bound,
and this is where the low-rate lower bound in Prop.~\ref{prop:pk_lr_lb} helps.

\begin{proposition}
In the PPNF, we have 
$ \E[N] \geq \sum_{k=1}^{\lfloor 1/\theta \rfloor} \underline{p}_k^{\textnormal{LR}}$.\label{prop:ENlb_lr}
\end{proposition}

A rigorous upper bound can be derived similarly but with more caution as
we cannot simply discard a number of terms in the sum.
The following lemma presents a bound based on Prop.~\ref{prop:pk_c_ub}.

\begin{proposition}
In the PPNF, $\E[N]$ is upper bounded by
\begin{equation*}
	\frac{e^{1+K}}{\sqrt{2\pi}}\frac{(cK)^{1-K}}{cK-1}
	+\frac{e}{c} (1+c)^{1-K}
	+\sum_{k=1}^{K-1} \bar\theta^{-\frac{\beta}{2}k(k-1)} \Delta_2(k),
\end{equation*}
for all $K\in \mathbb{N}\cap [e/c,\infty)$, where $\bar{\theta}=\max\{\theta,1\}$.
\label{prop:ENub}
\end{proposition}

The proof Prop.~\ref{prop:ENub} is based on upper bounding the tail terms
of the infinite sum and can be found in App.~\ref{app:ENATPf}.
Likewise, we can build a SMUD upper bound based on Prop.~\ref{prop:altbounds_pk} as follows.

\begin{proposition}[SMUD upper bound]
\label{prop:ENbound_alt}
The mean number of decodable users is upper bounded by
\begin{equation*}
 \E N \leq \sum_{k=1}^{K-1} \left(\frac{C(k)}{\Gamma(1-\beta)}\right)^k
		\frac{1}{\Gamma(1+k\beta)} 
		+ \frac{1}{\Gamma(1+K\beta)}\left(\frac{C(K)}{\Gamma(1-\beta)}\right)^K 
		\frac{\Gamma(1-\beta)}{\Gamma(1-\beta)-C(K)}, 
\end{equation*}
where $C(k)\triangleq \theta^{-\beta} \bar\theta^{-\frac{\beta}{2}(k-1)}$.
\end{proposition}

The idea of the proof Prop.~\ref{prop:ENbound_alt} closely resembles that of Prop.~\ref{prop:ENub}
and is thus omitted from the paper.

\begin{figure}[t]
\centering
\includegraphics[width=0.5\linewidth]{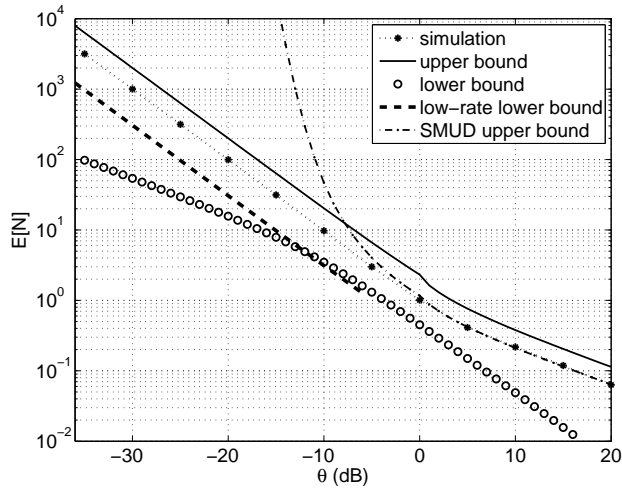}
\caption{The mean number of users that can be successively decoded in a 2-d uniform network
with path loss exponent $\alpha=4$.
Here, the upper bound, lower bound, low-rate lower bound, SMUD upper bound refer
to the bounds in Props.~\ref{prop:ENlb},~\ref{prop:ENub},~\ref{prop:ENlb_lr} and \ref{prop:ENbound_alt}, respectively.}
\label{fig:ENSimBounds}
\end{figure}

Fig.~\ref{fig:ENSimBounds} compares the bounds provided in Props.~\ref{prop:ENlb}, \ref{prop:ENlb_lr}, \ref{prop:ENub} and \ref {prop:ENbound_alt}
with simulation results in a uniform 2-d network with $\alpha=4$.
Although the low-rate lower bound can be calculated for all $\theta<1$,
it is only meaningful when $\theta$ is so small that 
the lower bound in Prop.~\ref{prop:ENlb} fails to capture the rate
at which $\E N$ grows with decreasing $\theta$.
Thus, we only plot the low-rate lower bound for $\theta<-5 \textnormal{\;dB}$.

As is shown in the figure, $\E N$ increases unboundedly with the decreasing of $\theta$,
which further confirms that SIC is particularly beneficial for low-rate applications
in wireless networks, such as node discovery, cell search, \emph{etc}.

Fig.~\ref{fig:ENSimBounds} also shows the different merits of the different closed-form bounds
presented above.
The bounds of Props.~\ref{prop:ENlb}~and~\ref{prop:ENub}
behave well in most of the regime where the practical systems operate.
In the lower SIR regime, \ie when $\theta\to 0$,
the low-rate lower bound outperforms the lower bound in Prop.~\ref{prop:ENlb}
which does not capture the asymptotic behavior of $\E N$.
The SMUD bound in Prop.~\ref{prop:ENbound_alt} provides
a tighter alternative to the upper bound in Prop.~\ref{prop:ENub} 
and is especially tight for $\theta>1$.

\subsection{The Aggregate Throughput\label{sec:AggThroughput}}

Although a smaller $\theta$ results in more effective SIC,
it also means the information rate at each transmitter is smaller.
Thus, it is interesting to see how the aggregate throughput defined in (\ref{equ:aggInfoRate})
changes with respect to $\theta$.
One way of estimate the aggregate throughput is by
using Props.~\ref{prop:ENlb}, \ref{prop:ENlb_lr}, \ref{prop:ENub} and \ref{prop:ENbound_alt}.

\begin{figure}
\centering
\includegraphics[width=0.5\linewidth]{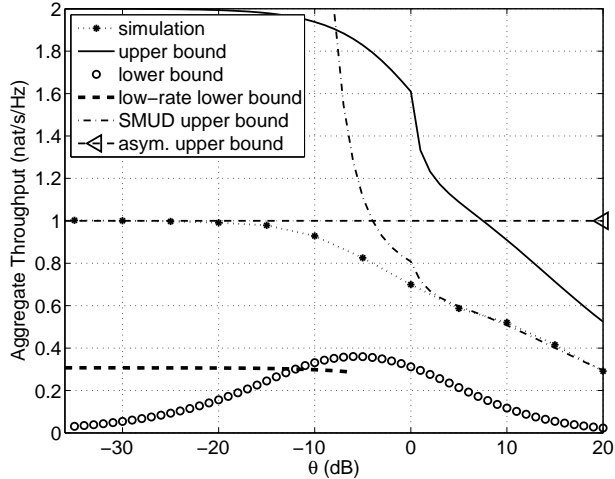}
\caption{Aggregate throughput at $o$ in a 2-d uniform network with
with path loss exponent $\alpha=4$, \emph{i.e.,} $\beta=\delta=2/\alpha=1/2$.
The upper bound, lower bound, low-rate lower bound and SMUD upper bound come from 
Props.~\ref{prop:ENub},~\ref{prop:ENlb},~\ref{prop:ENlb_lr}~and~\ref{prop:ENbound_alt} respectively.
In this case, the asymptotic upper bound (Prop.~\ref{prop:asymtightBD}) is $1/\beta-1=1\;$nats/s/Hz and is plotted with dashed line labeled by
a left-pointing triangle.
}
\label{fig:InfoRateSimBounds}
\end{figure}

Fig.~\ref{fig:InfoRateSimBounds} shows the total information rate as a function
of $\theta$ with analytical bounds and simulation.
Again, we only show the low-rate lower bounds for $\theta<-5\textnormal{\;dB}$.
In this case, we see that the lower bound
of the aggregate throughput becomes a non-zero constant when $\theta \rightarrow 0$
just like the upper bound.
Therefore, our results indicate that while the aggregate throughput
diminishes when $\theta\rightarrow \infty$,
it converges to a finite non-zero value when $\theta \rightarrow 0$.
In particular, by using Prop.~\ref{prop:ENub} and letting $\theta\to 0$, we can upper
bound the asymptotic aggregate throughput by $\frac{2}{\beta} -2$,
which turns out to be a loose bound.

Nevertheless,
it is possible to construct a better bound which improves (reduces) the bound
by a factor of $2$ and is numerically shown to be asymptotically tight (as is also shown in Fig.~\ref{fig:InfoRateSimBounds}
and will be proved below).
To show this better bound, we introduce the following lemma whose proof can be found in App.~\ref{app:ENATPf}. 

\begin{lemma}
The Laplace transform of $\xi_k I_k$ is
\begin{equation}
	\L_{\xi_k I_k}(s)=\frac{1}{(c(s)+1)^k},
\label{equ:Lxiklk}
\end{equation}
where $c(s)=s^\beta\gamma(1-\beta,s)-1+e^{-s}$.
\label{lem:LxikIk}
\end{lemma}

Then, we have the following asymptotic bound on the aggregate throughput.%

\begin{proposition}
The aggregate throughput $R=\log(1+\theta)\E[N]$ is (asymptotically) upper bounded by 
\begin{equation}
	\lim_{\theta\rightarrow 0} R \leq \frac{1}{\beta}-1.
\end{equation}
\label{prop:asymtightBD}
\end{proposition}
\begin{IEEEproof}
First, we have
\begin{align}
	\E[N]=\sum_{k=1}^\infty p_k \leq \sum_{k=1}^\infty \P(\xi_k I_k < 1/\theta)
				= \sum_{k=1}^\infty \int^{1/\theta}_0 f_{\xi_k I_k}(x) \d x
				= \int^{1/\theta}_0 \sum_{k=1}^\infty  f_{\xi_k I_k}(x) \d x. \label{equ:ENUpperBoundByfxiI}
\end{align}
In general, the RHS of (\ref{equ:ENUpperBoundByfxiI}) is not available in closed-form
since $f_{\xi_k I_k}$, the pdf of $\xi_k I_k$, is unknown.
However, when $\theta\rightarrow 0$, this quantity can be evaluated in the Laplace domain.
To see this, consider a sequence of functions $\left( f_n\right)_{n=1}^\infty$, where
$f_n(x)=\frac{1}{n}\sum_{k=1}^{n}f_{\xi_k I_k}(x),~\forall x>0$ and, obviously, 
$\int_0^\infty f_n(x) \d x = 1$ for all $n$. Thus,
\begin{equation}
\label{equ:ENequLaplace}
	1=\lim_{\theta\rightarrow 0} \frac{\int_0^{1/\theta} f_n(x) \d x}{\int_0^{\infty} e^{-\theta x} f_n(x) \d x}
	=\lim_{\theta\rightarrow 0} \frac{\int^{1/\theta}_0 \sum_{k=1}^\infty  f_{\xi_k I_k}(x) \d x}
	{\int_0^{\infty} e^{-\theta x} \sum_{k=1}^{\infty}    f_{\xi_k I_k}(x) \d x},~\forall n\in\mathbb{N}
\end{equation}
where
\begin{align*}
 \int_0^{\infty} e^{-\theta x} \sum_{k=1}^{\infty}    f_{\xi_k I_k}(x) \d x
 	= \sum_{k=1}^{\infty} \int_0^{\infty} e^{-\theta x} f_{\xi_k I_k}(x) \d x
	= \sum_{k=1}^{\infty} \L_{\xi_k I_k} (\theta).
\end{align*}
Comparing (\ref{equ:ENUpperBoundByfxiI}) and (\ref{equ:ENequLaplace}) yields that
\begin{equation*}
	\lim_{\theta\rightarrow 0} \frac{\E[N]}{\sum_{k=1}^{\infty} \L_{\xi_k I_k} (s)|_{s=\theta}} \leq 1,
\end{equation*} 
where $\L_{\xi_k I_k} (s)$ is given by Lemma~\ref{lem:LxikIk}.
Therefore, we have
\begin{align*}
	\lim_{\theta\rightarrow 0} \log(1+\theta) \E[N] &\leq \lim_{\theta\rightarrow 0} 
							\theta\sum_{k=1}^{\infty} \L_{\xi_k I_k} (\theta)		
										= \lim_{\theta\rightarrow 0} \frac{\theta}{c(\theta)}.
\end{align*}
The proof is completed by noticing that $\lim_{\theta\rightarrow 0} \frac{\theta}{c(\theta)}=\frac{1-\beta}{\beta}$.
\end{IEEEproof}

In the example considered in Fig.~\ref{fig:InfoRateSimBounds},
we see the bound in Prop.~\ref{prop:asymtightBD}
matches the simulation results.
Along with this example, we tested $\beta=1/3$ and $\beta=2/3$, and the bound is tight in both cases,
which is not surprising.
Because, in the proof of Prop.~\ref{prop:asymtightBD}, the only slackness introduced
is due to replacing $p_k$ with $\P(\xi_k^{-1}>\theta I_k)$,
and it is conceivable that, for every given $k$, 
this slackness diminishes in the limit,
since $\lim_{\theta\to 0 }\P(\xi_k^{-1}>\theta I_k)=\lim_{\theta\to 0}p_k=1$.
Thus, estimating $\E[N]$ by $\sum_{k=1}^\infty \P(\xi_k^{-1}>\theta I_k)$ is exact in the limit.

Many simulation results (including the one in Fig.~\ref{fig:InfoRateSimBounds}) suggest that
the aggregate throughput monotonically increases with decreasing $\theta$.
Assuming this is true, Prop.~\ref{prop:asymtightBD}
provides an upper bound on the aggregate throughput in the network for all $\theta$.
We also conjecture that this bound is asymptotically tight and thus can be achieved by driving the code rate
at every user to $0$, which is also backed by simulations (\emph{e.g.,} see Fig.~\ref{fig:InfoRateSimBounds}).

Since the upper bound is a monotonically decreasing function of $\beta$
we can design system parameters to maximize the achievable aggregate throughput provided that we can manipulate $\beta$ to some extent.
For example, since $\beta=\delta + b/\alpha$ and $\delta=d/\alpha$, one can try to reduce $b$ to increase the upper bound.
Note that $b$ is a part of the density function of the \emph{active} transmitters in the network and can be changed by independent thinning
of the transmitter process \cite{net:mh12}, and a smaller $b$ means the active transmitters are more clustered
around the receiver.
This shows that a MAC scheme that introduces clustering 
has the potential to achieve higher aggregate throughput
in the presence of SIC.

\subsection{A Laplace Transform-based Approximation\label{sec:LTApprox}}

Lemma~\ref{lem:LxikIk} gives the Laplace transform of $\xi_k I_k$,
which completely characterizes $\P(\xi_k^{-1}>\theta I_k)$, an important quantity
in bounding $p_k$, $\E [N]$ and thus $R$.
As analytically inverting (\ref{equ:Lxiklk}) seems hopeless,
a numerical inverse Laplace transform naturally becomes an alternative to provide more accurate
system performance estimates.
However, the inversion (numerical integration in complex domain) is generally
difficult to interpret and offers limited insights into the system performance.

On the other hand, $\L_{\xi_k I_k}(\theta)=\P(H>\theta \xi_k I_k)$
for an unit-mean exponential random variable $H$.
This suggests to use $\L_{\xi_k I_k}(\theta)$ to approximate $\P(\xi_k^{-1}>\theta I_k)$.
We would expect such an approximation to work for (at least) small $\theta$.
Because, first, it is obvious that for each $k$, this approximation is exact as $\theta\rightarrow 0$
since in that case both the probabilities go to 1;
second and more importantly, Prop.~\ref{prop:asymtightBD} shows that
the approximated $R$ based on this idea is asymptotically exact.

\begin{figure}[t]
\centering
\includegraphics[width=0.5\linewidth]{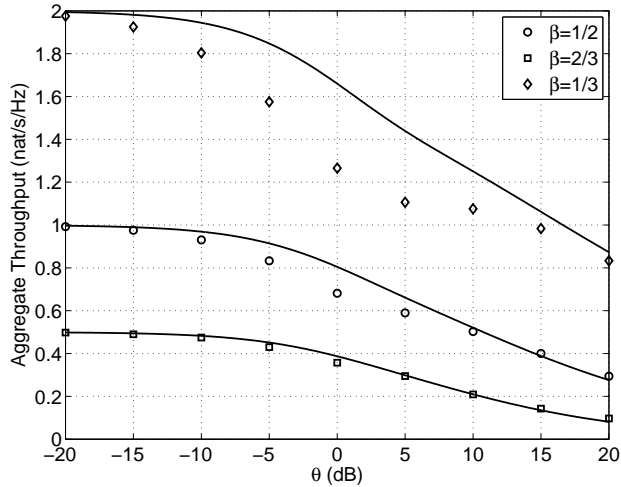}
\caption{Simulated and approximated, by (\ref{equ:LTapproxR}), aggregate throughput at $o$ in a 2-d uniform network. }
\label{fig:pkapprox}
\end{figure}

According to such an approximation, we have
\begin{equation}
	R\approx \frac{\log(1+\theta)}{c(\theta)}=\frac{\log(1+\theta)}{\theta^\beta\gamma(1-\beta,\theta)-1+e^{-\theta}}.
\label{equ:LTapproxR}
\end{equation}
This approximation is compared with simulation results in Fig.~\ref{fig:pkapprox},
where we consider $\beta={1}/{3}$, ${1}/{2}$ and ${2}/{3}$.
As shown in the figure, the approximation is tight
from -20\;dB to 20\;dB which covers the typical values of $\theta$.
Also, as expected, the approximation is most accurate in the small $\theta$ regime\footnote{
The fact that the approximation is also accurate for very large $\theta$ is more of a coincidence,
as the construction of the approximation ignores ordering requirement within the strongest (decodable) $k$ users
and is expected to be fairly inaccurate when $\theta\to \infty$ (see Lemma~\ref{lem:d-dim}).},
which is known to be the regime where SIC is most useful\cite{Zhang12globecom,net:Weber07tit,BlomerJindal09icc}.

\section{The Effect of Noise\label{sec:noise}}

In many wireless network outage analyses,
the consideration of noise is neglected mainly due to the argument that
most networks are interference-limited (without SIC).
However, this is not necessarily the case for a receiver with SIC capability,
especially when a large number of transmitters
are expected to be successively decoded.
Since the users to be decoded in the later stages have significantly weaker
signal power than the users decoded earlier,
even if for the first a few users interference dominates noise,
after decoding a number of users, the effect of noise can no longer be neglected.

Fortunately, most of the analytical bounds derived before can be adapted to the case where noise is considered.
If we let $\tilde{N}$ be the number of users that can be successively decoded in the presence of noise of power $W$,
we can define $p_k^W\triangleq \P(\tilde{N}\geq k)$ to be the probability of successively decoding
at least $k$ users in the presence of noise.
Considering the (ordered) PLPF $\Xi=\{\xi_i=\frac{\|x\|^\alpha}{h_x}\}$ as before, we can write $p_k^W$ as
\begin{equation*}
	p_k^W \triangleq \P\left(\xi_i^{-1}>\theta (I_i + W),\; \forall i\leq k\right),
\end{equation*}
and we have a set of analogous bounds as in the noiseless case.

\begin{lemma}
In a noisy PPNF, the probability of successively decoding $k$ users is bounded as follows:
\begin{itemize}
\item
$p_k^W \geq
(1+\theta)^{-\frac{\beta k(k-1)}{2}}\P(\xi_k^{-1}>\theta \left(I_{k}+W)\right)$
\item
$p_k^W \leq \theta^{-\frac{\beta k(k-1)}{2}}\P(\xi_k^{-1}>\theta (I_{k}+W))$
\end{itemize}
where $\Xi_\beta=\{\xi_i\}$ is the corresponding SPLPF and $I_{k} \triangleq \sum_{j=k+1}^\infty \xi_j^{-1}$.
\label{lem:d-dimw/noise}
\end{lemma}

\begin{IEEEproof}
The proof is analogous to the proof of Lemma~\ref{lem:d-dim} with two major distinctions:
First, we need to redefine the event $A_i$ to be $\{\xi_i^{-1}>\theta (I_{i}+W)\}$.
Second, Fact~\ref{fact:SPLPF} does not hold in the noisy case,
and thus the original PLPF (instead of the normalized SPLPF) needs to be considered.
However, fortunately, this does not introduce any difference on the order statistics of the first $k-1$ smallest elements in $\Xi$
conditioned on the $\xi_k$, and thus the proof can follow exactly the same as that of Lemma~\ref{lem:d-dim}.
\end{IEEEproof}

Thanks to Lemma~\ref{lem:d-dimw/noise}, bounding $p_k^W$ reduces to bounding $\P(\xi_k^{-1}>\theta \left(I_{k}+W)\right)$.
Ideally, we can bound $\P(\xi_k^{-1}>\theta \left(I_{k}+W)\right)$ by reusing the bounds we have on $\P(\xi_k^{-1}>\theta I_k)$.
Yet, this method does not yield closed-form expressions (in most cases such bounds will be in an infinite integral form).
Thus, we turn to a very simple bound which can still illustrate
the distinction between the noisy case and the noiseless case.

\begin{lemma}
In a noisy PPNF, we have
\begin{equation}
	\P(\xi_k^{-1}>\theta \left(I_{k}+W)\right)\leq \bar\gamma(k,\frac{\bar{a}}{\theta^\beta W^\beta}),
\end{equation}
where $\bar{a}=a\delta c_d \E[h^\beta]/\beta$, $\beta=\delta+b/\alpha$, and $\delta=d/\alpha$.
\label{lem:SimpleNoisyUpperBound}
\end{lemma}
\begin{IEEEproof}
First, note that $\P(\xi_k^{-1}>\theta \left(I_{k}+W)\right)\leq \P(\xi_k<\frac{1}{\theta W})$ which equals the probability
that there are no fewer than $k$ elements of the PLPF smaller than $1/\theta W$.
By Lemma~\ref{lem:mapping}, the number of elements of the PLPF in $(0,1/\theta W)$ is Poisson distributed
with mean $\bar{a}/\theta^\beta W^\beta$, and the lemma follows.
\end{IEEEproof}

Although being a very simple bound, Lemma~\ref{lem:SimpleNoisyUpperBound} directly leads to the following proposition
which contrasts what we observed in the noiseless network.

\begin{proposition}
In a noisy PPNF, the aggregate throughput goes to 0 as $\theta\rightarrow 0$.
\label{prop:R->zero}
\end{proposition} 
\begin{IEEEproof}
Combining Lemma~\ref{lem:d-dimw/noise} and Lemma~\ref{lem:SimpleNoisyUpperBound},
we have
\begin{align*}
	\E [{N}]	&=\sum_{k=1}^{\infty}p_k^W			
			\leq \sum_{k=1}^{\infty}\P(\xi_k^{-1}>\theta \left(I_{k}+W)\right)		
			\leq \sum_{k=1}^{\infty} \bar\gamma(k,\frac{\bar{a}}{\theta^\beta W^\beta})
			=\bar{a}/\theta^\beta W^\beta.
\end{align*}
In other words, $\E [N]$ is upper bounded by the mean number of elements of the PLPF in $(0,1/\theta W)$.
Then, it is straightforward to show that $\lim_{\theta \rightarrow 0} R \leq \lim_{\theta\rightarrow 0} \bar{a}\theta^{1-\beta}/W^{\beta}$,
and the RHS equals zero since $\beta\in(0,1)$.
\end{IEEEproof}

Since it is obvious that $\lim_{\theta\to\infty}R=0$,
we immediately obtain the following corollary.

\begin{corollary}
There exists at least one optimal $\theta>0$
that maximizes the aggregate throughput in a noisy PPNF.
\label{cor:exist:optimal_theta}
\end{corollary}

\begin{figure}[t]
\centering
\includegraphics[width=0.5\linewidth]{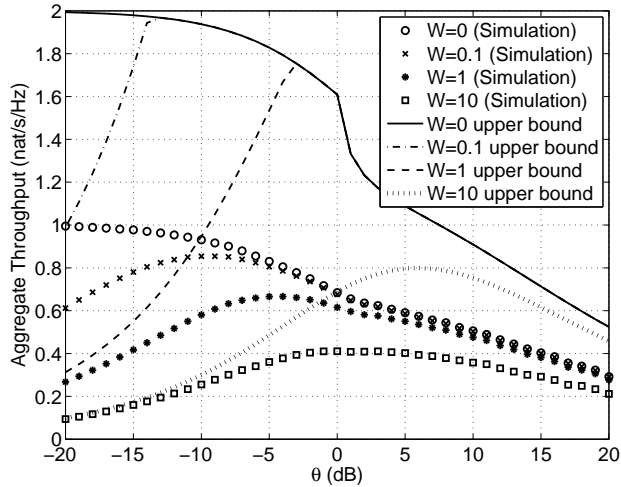}
\caption{Aggregate throughput at $o$ in a 2-d uniform network with noise.
Here, the path loss exponent $\alpha=4$. Three levels of noise are considered: $W=0.1$, $W=1$ and $W=10$.}
\label{fig:InfoRateSINR}
\end{figure}

As is shown in the proof of Prop.~\ref{prop:R->zero},
$\bar{a}/\theta^\beta W^\beta$ is an upper bound on $\E[N]$.
We can obtain an upper bound on the aggregate throughput
by taking the minimum of $\log(1+\theta) \bar{a}/\theta^\beta W^\beta$
and the upper bound shown in Fig.~\ref{fig:InfoRateSimBounds}.
Fig.~\ref{fig:InfoRateSINR} compares the upper bounds with simulation results,
considering different noise power levels.
This figure shows that the noisy bound becomes tighter and
the interference bound becomes looser as $\theta\to 0$.
This is because as $\theta$ decreases the receiver is expected to successively
decode a larger number of users.
The large amount of interference canceled makes the residual interference
(and thus the aggregate throughput)
dominated by noise.
In this sense, the optimal per-user rate mentioned in Cor.~\ref{cor:exist:optimal_theta}
provides the right \emph{balance} between interference and noise
in noisy networks.

Thanks to Lemma~\ref{lem:SimpleNoisyUpperBound},
we see that Prop.~\ref{prop:scale-invariant} clearly does not hold for noisy networks.
Nevertheless,
there is still a monotonicity property in noisy networks,
analogous to the scale-invariance property in noiseless networks, 
as stated by the following proposition.

\begin{proposition}[Scale-monotonicity]
\label{prop:scale-monotonicity}
For two PLPF $\Xi$ and $\bar{\Xi}$ with intensity measure $\Lambda([0,r])=a_1 r^\beta$ and $\mu([0,r])=a_2 r^\beta$,
where $a_1$ and $a_2$ are positive real numbers and $a_1\leq a_2$,
we have $p_k^W(\Xi)\leq p_k^W(\bar{\Xi}),~\forall k\in\mathbb{N}$.
\end{proposition}
\begin{IEEEproof}
See App.~\ref{app:ENATPf}.
\end{IEEEproof}

Combining Lemma~\ref{lem:mapping} and Prop.~\ref{prop:scale-monotonicity} yields
the following corollary, since $\E[h^\beta]\leq 1$ given that $\E[h]=1$ (recall that $\beta\in(0,1)$).

\begin{corollary}
In a noisy PPNF, fading reduces $p_k^W$, the mean number of users that can be successively decoded,
and the aggregate throughput.
\label{cor:fadingIsBad}
\end{corollary}

Since random power control, \emph{i.e.,} randomly varying the transmit power at each transmitter
under some mean and peak power constraint \cite{net:Zhang12tcom,net:Zhang12twc},
can be viewed as a way of manipulating the fading distribution,
Cor.~\ref{cor:fadingIsBad} also indicates that (iid) random power control
cannot increase the network throughput in a noisy PPNF.

\section{Application In Heterogeneous Cellular Networks\label{sec:HetNet}}

\subsection{Introduction}

The results we derived in the previous sections apply to many types of wireless networks.
One of the important examples is the downlink of heterogeneous cellular networks (HCNs).
HCNs are multi-tier cellular networks where the marcro-cell base stations (BSs) are overlaid
with low power nodes such as pico-cell BSs or femto-cell BSs.
This heterogeneous architecture is believed to be part of the solution package
to the exponentially growing data demand of cellular users \cite{GhoAnd12,Madan10JSAC}.
However, along with the huge cell splitting gain and deployment flexibility,
HCNs come with the concern that the increasing interference may diminish or
even negate the gain promised by cell densification.
This concern is especially plausible when some of the tiers in the network can
have closed subscriber groups (CSG), \ie some BSs only serve a subset of the users
and act as pure interferers to other users.

There are multiple ways of dealing with the interference issues in HCNs
including exploiting MIMO techniques \cite{HuangAndrewsGuoHeathBerry12tit,VazeHeath12},
coordinated multi-point processing (CoMP) \cite{IrmerJungnickel2011CommMag,DLeeSayans2012CommMag} and 
inter-cell interference coordination (ICIC) \cite{EICIC2011TWC,BarbieriDamnjanovicHorn2012TWC,Zhang14twc}.
In addition, successive interference cancellation is also believed to play
an important part in dealing with the interference issues in HCNs \cite{DamnjanovicMalladi2011WC}.

In this section, leveraging tools developed in the previous sections,
we will analyze the potential benefit of SIC in ameliorating the interference within and across tiers.
The key difference between the analysis in this section and those in Section~\ref{sec:EN}
is that
in HCNs, the receiver (UE) is only interested in being connected to
\emph{one} of the transmitters (BSs) %
whereas in Section~\ref{sec:EN}, we assumed that the receiver is interested in the message
transmitted from all of the transmitters.

\begin{figure}[t]
\centering
\includegraphics[width=0.4\linewidth]{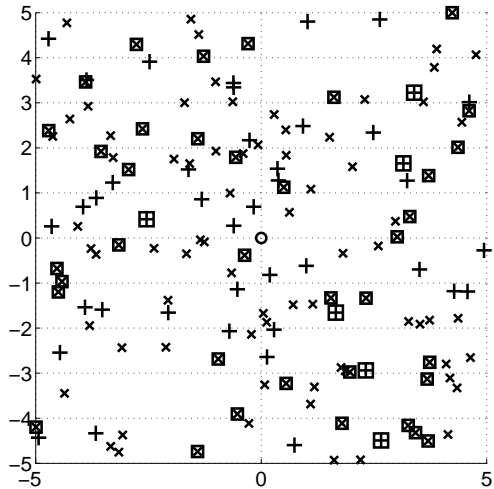}
\caption{A 2-tier HCN with 10\% of Tier 1 (macrocell) BSs (denoted by $\sf +$) overloaded
and 30\% of Tier 2 (femtocell) BSs (denoted by $\times$) configured as closed.
A box is put on the BS whenever it is non-accessible (\emph{i.e.,} either configured
as closed or overloaded). The $\sf o$ at origin is a typical receiver.}
\label{fig:HetNetIll}
\end{figure}

We model the base stations (BSs) in a $K$-tier HCN
by a family of marked Poisson point processes (PPP) $\{\hat{\Phi}_i,~i\in[K]\}$,
where $\hat{\Phi}_i=\{(x_j,h^{(i)}_{x_j},t^{(i)}_{x_j})\}$ represents the BSs of the $i$-th tier, 
$\Phi_i = \{x_j\} \subset \R^2$ are uniform\footnote{Although we only consider uniformly distributed BSs in this section, 
with the results in previous sections, generalizing the results to non-uniform (power-law density) HCNs is straightforward.} PPPs with intensity $\lambda_i$,
$h^{(i)}_x$ is the iid (subject to distribution $f^{(i)}_h(\cdot)$)
fading coefficient of the link from $x$ to $o$,
and $t^{(i)}_x$ is the \emph{type} of the BS and is an iid Bernoulli random variable with
$\P(t^{(i)}_x=1)=\pi^{(i)}$ and $\P(t^{(i)}_x=0)=1-\pi^{(i)}$.
If $t^{(i)}_x=1$, we call the BS $x$ \emph{accessible} and otherwise \emph{non-accessible}.
Using $t_x$ to model the accessibility of the BSs enables us the incorporate the effect
of some BS being configured with CSG and thus acts as pure interferers to the typical UE.\footnote{
In addition to modeling the CSG BSs, the non-accessible BSs can also be interpreted as
overloaded/biased BSs \cite{GhoAnd12} or simply interferers outside the cellular system.}
For a typical receiver (UE) at $o$, the received power from BS $x\in\Phi_i$
is $P^{(i)}h^{(i)}_x \|x\|^{-\alpha}$,
where $P^{(i)}$ is the transmit power at BSs of tier $i$,
and $\alpha$ is the path loss exponent.
Also note that since this section focuses on 2-d uniform networks,
we have $\beta=2/\alpha$.
An example of a two tier HCN is shown in Fig.~\ref{fig:HetNetIll}.

An important quantity that will simplify our analysis in
the $K$-tier HCN is the \emph{equivalent access probability} (EAP)
defined as below.

\begin{definition}
Let
\begin{equation*}
	Z \triangleq \sum_{i=1}^K\lambda_i \E[(h^{(i)})^\beta](P^{(i)})^\beta.
\end{equation*}
The \emph{equivalent access probability} (EAP) is the following weighted average of the
individual access probabilities $\pi^{(i)}$:
\begin{equation*}
	\eta = \frac{1}{Z}\sum_{i=1}^K \pi^{(i)} \lambda_i \E[(h^{(i)})^\beta](P^{(i)})^\beta.
\end{equation*}
\end{definition}

Thanks to the obvious similarity between this HCN model and our PPNF model introduced in
Section~\ref{sec:sysmodel}, we can define the \emph{marked} PLPF as follows.

\begin{definition}
The \emph{marked PLPF} corresponding to the tier $i$ network is
$\hat{\Xi}_i=\{(\frac{\|x\|^\alpha}{h_x P^{(i)}},t_x):x\in\Phi_i\}$,
with $\Xi_i \triangleq \{\frac{\|x\|^\alpha}{h_x P^{(i)}}:x\in\Phi_i\}$
being the (ground) PLPF.
\end{definition}

Furthermore, we denote the union of the $K$ marked PLPFs and ground PLPFs 
as $\hat{\Xi}\triangleq \bigcup_{i=1}^K \hat{\Xi}_i$ and
$\Xi\triangleq\bigcup_{i=1}^K \Xi_i$, respectively.
Then, we have the following lemma.

\begin{lemma}
The PLPF corresponding to the $K$-tier heterogeneous cellular BSs is a marked inhomogeneous PPP
$\hat{\Xi}=\{(\xi_i,t_{\xi_i})\}\subset \R^+\times \{0,1\}$,
where the intensity measure of $\Xi=\{\xi_j\}$ is $\Lambda([0,r])=Z \pi r^\beta$
and the marks $t_\xi$ are iid Bernoulli with
$\P(t_\xi=1)=\eta$.
\label{lem:heteroPLPF}
\end{lemma}

Based on the mapping theorem, the independence between $t^{i}_x$ and the fact that the superposition of PPPs is still a PPP,
the proof of Lemma~\ref{lem:heteroPLPF} is straightforward and thus omitted from the paper.
Despite the simplicity of the proof, the implication of Lemma~\ref{lem:heteroPLPF} is
significant:
the effect of
the different transmit powers, fading distributions and access probabilities of the $K$-tiers
of the HCN can all be subsumed by the two parameters $Z$ and $\eta$.

\subsection{The Coverage Probability\label{subsec:theCovProb}}

An important quantity in the analysis of the downlink of heterogeneous cellular networks is
the coverage probability, which is defined as the probability of a typical UE successfully
connecting to (at least) one of the accessible BSs
(after possibly canceling some of the non-accessible BSs).

\subsubsection{Without SIC}

Using the PLPF framework we established above and assuming that the UE \emph{cannot} do SIC and the system is interference-limited,
we can simplify the coverage probability in the 
$K$-tier cellular network to
\begin{equation}
	P_c = \P\left(\frac{\xi_\ast^{-1}}{\sum_{\xi\in\Xi\backslash\{\xi_\ast\}}\xi^{-1}}>\theta \right),
\label{equ:PcwithoutSIC}
\end{equation}
where $\xi_\ast\triangleq \argmax_{\xi\in\Xi}t_\xi\xi^{-1}$, and $\theta$ is the SIR threshold.

Note that the coverage probability in (\ref{equ:PcwithoutSIC})
does not yield a closed-form expression in general \cite{DhiGanC2011a}.
However, for $\theta\geq 1$, we can deduce
\begin{equation}
	P_c = \eta p_1 = \frac{\eta \, {\sinc\beta}}{{\theta^\beta}},
	\label{equ:Pcanalytical}
\end{equation}
by combining Cor.~\ref{cor:coverageProb_highSIR}
with the fact that the marks $\{t_i\}$ are independent from $\Xi$.
More precisely, when $\theta\geq 1$ (SMUD Regime), it is not possible for the UE to decode
any BS other than the strongest BS without SIC\footnote{Intuitively,
decoding any BS weaker than the strongest BS implies that
this BS is stronger than the strongest BS and causes contradiction. This argument can be made
rigorous by applying Lemma~\ref{lem:k-strongest} (in App.~\ref{app:Pxik>thetaIk}) for the case $k=1$.}.
Thus, the coverage probability without SIC is the product of the probability that the strongest BS being
accessible $\eta$ and the probability of decoding the strongest BS $p_1$.

\subsubsection{With SIC}

Similar to (\ref{equ:PcwithoutSIC}), we can define the coverage probability when the UE
has SIC capability. In particular, the coverage probability $P_c^\textnormal{SIC}$ is the probability that
after canceling a number of non-accessible BSs, the signal to (residual) interference
ratio from the any of the accessible BSs is above $\theta$.
Formally, with the help of the PLPF,
we define the following event of coverage which happens with probability $P_c^\textnormal{SIC}$.

\begin{definition}[Coverage with (infinite) SIC capability]
A UE with infinite SIC capability is \emph{covered}
iff there exists $l\in \mathbb{N}$ and $k\in\{i: t_i=1\}$
such that $\xi_i^{-1}>\theta I_i,\;\forall i\leq l$
and $\xi_k^{-1}>\theta I^{!k}_l$,
where $I^{!k}_{l}\triangleq \sum_{j\geq l+1}^{j\neq k} \xi_j^{-1}$.
\label{def:coverage}
\end{definition}

In words, Def.~\ref{def:coverage} says that the UE is covered if and only if
there exists an integer pair $(k,l)$, such that the $k$-th strongest BS is accessible
and can be decoded after successively canceling $l$ BSs.

With the help of PLPF and the parameters we defined in the analysis of the PPNF,
the following lemma describes this probability in a neat formula.

\begin{proposition}
In the $K$-tier heterogeneous cellular network, 
the coverage probability of a typical UE with SIC is
\begin{equation*}
	P_c^\textnormal{SIC}=\sum_{k=1}^\infty (1-\eta)^{k-1} \eta p_k,
\end{equation*}
where $p_k=p_k(\Xi)$ is the probability of successively decoding
at least $k$ users in a PLPF on $\R^+$ with intensity measure
$\Lambda([0,r])=Z \pi r^\beta $.
\label{prop:UESIC}
\end{proposition}

\begin{IEEEproof}
See App.~\ref{app:HCNproofs}.
\end{IEEEproof}

Thanks to Prop.~\ref{prop:UESIC} we can quantify the coverage probability
of the HCN downlink using the bounds on $p_k$ we obtained in Section~\ref{sec:boundsonpk}.
In particular, based on Prop.~\ref{prop:pk_hr_lb}, a lower bound can be found as
\begin{equation}
	P_{c}^\textnormal{SIC}\geq \sum_{k=1}^{K} (1-\eta)^{k-1} \eta (1+\theta)^{-\frac{\beta k(k-1)}{2}}\Delta_1(k),
	\label{equ:PcSIC_lb}
\end{equation}
where the choice of $K$ affects the tightness of the bound.
Although a rigorous upper bound cannot be obtained by simply discarding some terms from the sum,
we can easily upper bound the tail terms of it.
For example, based on Prop.~\ref{prop:pk_c_ub} we have 
\begin{equation}
	P_{c}^\textnormal{SIC}\leq \sum_{k=1}^{K} (1-\eta)^{k-1} \eta \bar{\theta}^{-\frac{\beta}{2}k(k-1)}  \Delta_2(k)
		+(1-\eta)^{K+1},
		\label{equ:PcSIC_ub}
\end{equation}
where $\bar{\theta}=\max\{\theta,1\}$ and $(1-\eta)^{K+1}$ bounds the residual terms in the infinite sum.
Likewise, the SMUD upper bound on $p_k$ in Section~\ref{subsec:AlternativeBounds} leads to

\begin{multline}
 P^{\textnormal{SIC}}_c \leq \frac{\eta}{1-\eta}\sum_{k=1}^{K-1} \left(\frac{(1-\eta)C(k)}{\Gamma(1-\beta)}\right)^k
		\frac{1}{\Gamma(1+k\beta)} \\
		+ \frac{\eta}{1-\eta}\frac{1}{\Gamma(1+K\beta)}\left(\frac{(1-\eta)C(K)}{\Gamma(1-\beta)}\right)^K 
		\frac{\Gamma(1-\beta)}{\Gamma(1-\beta)-(1-\eta)C(K)}  , 
		\label{equ:PSICc_alt_ub}
\end{multline}
where $C(k)\triangleq \theta^{-\beta} \bar\theta^{-\frac{\beta}{2}(k-1)}$.

Besides these bounds, we can also use the approximation established in Section~\ref{sec:LTApprox}
to obtain an approximation on the coverage probability in closed-form.
In particular, we had
\begin{equation*}
	p_k\approx \L_{\xi_k I_k}(s)|_{s=\theta} = \frac{1}{\left(c(\theta)+1\right)^k},
\end{equation*}
where $c(\theta)=\theta^\beta\gamma(1-\beta,\theta)-1+e^{-\theta}$.
Combing this with Prop.~\ref{prop:UESIC}, we have
\begin{equation}
	P_{c}^\textnormal{SIC}\approx \frac{\eta}{1-\eta} \sum_{k=1}^{\infty} \left(\frac{1-\eta}{1+c(\theta)}\right)^k
						=\frac{\eta}{\eta+c(\theta)}.
						\label{equ:PcSIC_LTA}
\end{equation}

\begin{figure}[t]
\centering
\includegraphics[width=0.5\linewidth]{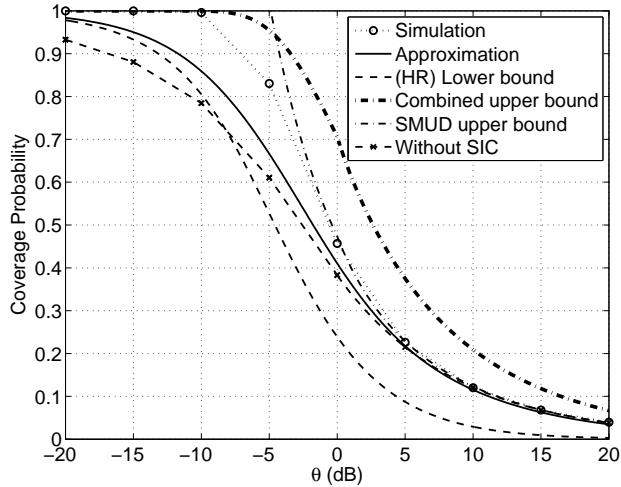}
\caption{The coverage probability (with infinite SIC capability) as a function of SIR threshold $\theta$ in HCNs with $\eta=0.6$ and $\alpha=4$.
The (Laplace-transform-based) approximation, high-rate lower bound, combined upper bounds of $P_c^\textnormal{SIC}$ and SMUD upper bound
is calculated according to (\ref{equ:PcSIC_LTA}), (\ref{equ:PcSIC_lb}), (\ref{equ:PcSIC_ub}) and \eqref{equ:PSICc_alt_ub}, respectively.
The coverage probability in the case without SIC (a problem also studied in \cite{net:Dhillon12jsac,DhiGanC2011a}) is also plotted for comparison,
where the $\theta\geq 0\textnormal{\;dB}$ part is analytically obtained by
(\ref{equ:Pcanalytical}) and the $\theta< 0\textnormal{\;dB}$ part
is based on simulation.
}
\label{fig:PcSIC_wrt_theta}
\end{figure}

In Fig.~\ref{fig:PcSIC_wrt_theta}, we compare these bounds and the approximation
with simulation results.
These bounds give reasonably good estimates on the coverage probability
throughout the full range of the SIR threshold $\theta$. 
In comparison with the coverage probability when no SIC is available,
we see that a significant gain can be achieved by SIC when the SIR threshold $\theta$
is between $-10$\;dB and $-5$\;dB.
This conclusion is, of course, affected by $\eta$.
The effect of $\eta$ will be further explored in Section~\ref{subsec:FiniteSIC}.

\subsection{The Effect of the Path Loss Exponent $\alpha$}

When $\theta\geq 1$, we can also lower bound the coverage probability
using the SMUD bound in Prop. \ref{prop:altbounds_pk}, which leads to
\begin{equation}
	P^{\textnormal{SIC}}_c\geq  \frac{\eta}{1-\eta}\sum_{k=1}^K  
						\frac{1}{(1+\theta)^{\frac{\beta}{2}k(k-1)}\Gamma(1+k\beta)}\left(\frac{1-\eta}{\theta^\beta\Gamma(1-\beta)}\right)^k						,\; \forall K\geq 1,
						\label{equ:PSICc_alt_lb}
\end{equation}
where we take a finite sum in the place of an infinite one. 
The error term associated
with this approximation is upper bounded as
\begin{equation}
	\sum_{k=K+1}^\infty  
						\frac{1}{(1+\theta)^{\frac{\beta}{2}k(k-1)}\Gamma(1+k\beta)}\left(\frac{1-\eta}{\theta^\beta\Gamma(1-\beta)}\right)^k
						\leq \frac{1}{\Gamma(1+(K+1)\beta)}\frac{C_2^{K+1}}{1-C_2},
						\label{equ:Psic_lb_errub_beta}
\end{equation}
where $C_2 = \frac{1-\eta}{(1+\theta)^{\frac{\beta}{2}K}\theta^\beta \Gamma(1-\beta)}$.
Since (\ref{equ:Psic_lb_errub_beta})
decays super-exponentially with $K$, a small $K$ typically ends up with a quite accurate estimate.

\begin{figure}[t]
\centering
\includegraphics[width=0.5\linewidth]{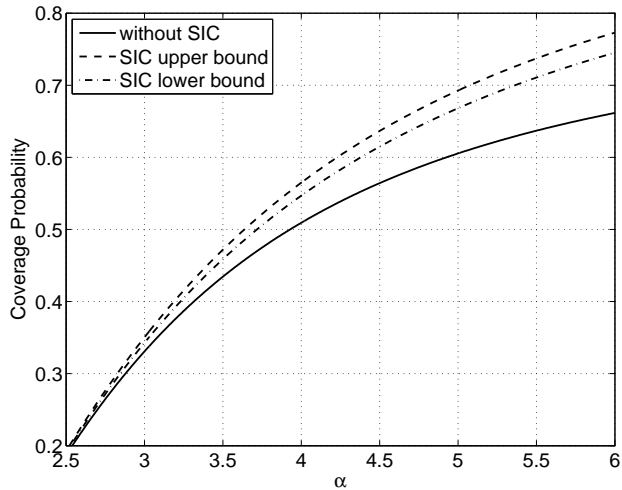}
\caption{Comparison between coverage probability with and without SIC in HCNs with $\eta=0.8$,
$\theta=1$.
Here, the upper and lower bounds are based on (\ref{equ:PSICc_alt_ub})
and (\ref{equ:PSICc_alt_lb}), respectively.}
\label{fig:PcSIC_wrt_alpha}
\end{figure}

Fig.~\ref{fig:PcSIC_wrt_alpha} plots the coverage probability as a function of the path loss
exponent $\alpha$. Here,
the coverage probability without SIC $P_c^\textnormal{SIC}$ is
given by (\ref{equ:Pcanalytical}).
The figure shows that the absolute gain of coverage probability due to SIC
is larger for larger path loss exponent $\alpha$.
Although our model here does not explicitly consider BS clustering,
by the construction of the PLPF in Section~\ref{sec:PLPF},
we can expect a larger gain due to SIC for clustered BSs.
Further numerical results also show that the gain is larger when $\eta$ is smaller,
\emph{i.e.,} there are more non-accessible BSs.

\subsection{Average Throughput}

\begin{figure}[t]
\centering
\includegraphics[width=0.5\linewidth]{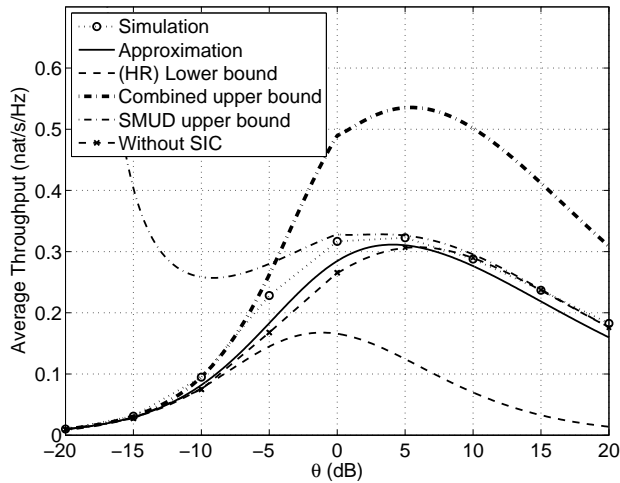}
\caption{The average throughput as a function of SIR threshold $\theta$ in HCNs with $\eta=0.6$ and $\alpha=4$.
The (Laplace-transform-based) approximation, lower bound and upper bounds of $P_c^\textnormal{SIC}$ 
is calculated according to (\ref{equ:PcSIC_LTA}), (\ref{equ:PcSIC_lb}), (\ref{equ:PcSIC_ub}) and \eqref{equ:PSICc_alt_ub}, respectively.
The non-outage throughput in the case without SIC is plotted for comparison,
where the $\theta\geq 0\textnormal{\;dB}$ part is analytically obtained by
(\ref{equ:Pcanalytical}) and the $\theta< 0\textnormal{\;dB}$ part
is based on simulation.
}
\label{fig:AvgThroughput_wrt_theta}
\end{figure}

Reducing the SIR threshold $\theta$ decreases the throughput of the
UE under coverage.
Similar to our analyses to the aggregate throughput, we can define the average
throughput as
\begin{equation*}
		T\triangleq \log(1+\theta) P_{c}^\textnormal{SIC}.
\end{equation*}
For the case without SIC, the definition is simplified as $T\triangleq \log(1+\theta) P_{c}$.
The average throughput is different from the aggregate throughput defined in Section~\ref{sec:AggThroughput}
in that we do not allow multiple packet reception in this case.

Fig.~\ref{fig:AvgThroughput_wrt_theta} shows how the average throughput change as a function
of $\theta$ with the same set of parameters as in Fig.~\ref{fig:PcSIC_wrt_theta}.
Comparing these two figures, we find that while SIC is particularly useful in terms of
coverage in combination with low-rate codes (low $\theta$),
the usefulness of SIC in terms of average throughput can be marginal.
For this particular set of parameters,
the average throughput is maximized at $\theta$ about $5$\;dB,
a regime where SIC is not very useful.
On the positive side, as we will show in Section~\ref{subsec:FiniteSIC},
for such $\theta$,
most of the gain of SIC can be obtained by simply canceling a very
small number of non-accessible BSs.

\subsection{Finite SIC Capabilty\label{subsec:FiniteSIC}}

In real cellular network settings, 
the assumption that the UEs have the ability to successively
decode an infinite number of interferers is impractical and conceivably
unnecessary in achieving the coverage gain.
Thus, it is important to evaluate the performance gain of SIC when the UEs
have only a limited ability of interference cancellation.
Since the latency is likely to be the most critical factor in practical
systems, we consider the case where the UE can cancel at most $n-1$ interferers.
Formally, we define the event of coverage for a UE with $n$-layer SIC capability as follows.

\begin{definition}[Coverage with $n$-layer SIC capability]
A UE with $n$-layer SIC capability is \emph{covered}
iff there exists $l\in[n-1]$ and $k\in\{i: t_i=1\}$
such that $\xi_i^{-1}>\theta I_i,\;\forall i< l$
and $\xi_k^{-1}>\theta I^{!k}_l$.
\label{def:coverageFinite}
\end{definition}

Comparing Def.~\ref{def:coverage} with Def.~\ref{def:coverageFinite},
we see that the only difference is that the integer pair $(k,l)$ has to satisfy $l\leq n-1$
which enforces the finite SIC capability constraint.
We will use $P^{\textnormal{SIC}}_{c,n}$ to denote
the coverage probability for a typical UE with $n$-layer SIC capability.
As two special cases, we have $P^{\textnormal{SIC}}_{c,1}=P_c$ and 
$P^{\textnormal{SIC}}_{c,\infty} = P^{\textnormal{SIC}}_{c}$.

Following a similar procedure in the proof of Prop.~\ref{prop:UESIC},
we find a lower bound on $P_{c,n}^\textnormal{SIC}$ which is exact when $\theta\geq 1$.

\begin{proposition}
In the $K$-tier heterogeneous cellular network, 
the coverage probability of a typical UE with $n$-layer SIC capability is
\begin{equation}
\label{equ:P_c,n}
	P_{c,n}^\textnormal{SIC}\geq \sum_{k=1}^{n} (1-\eta)^{k-1} \eta p_k,
\end{equation}
where the equality holds when $\theta\geq 1$.
\label{prop:UESICn}
\end{proposition}

the proof of Prop.~\ref{prop:UESICn} is analogous to that of Prop.~\ref{prop:UESIC} with extra care
about the possibility that an accessible BS can be decoded without canceling \emph{all}
the nonaccessible BSs stronger than it.
Details of proof can be found in App.~\ref{app:HCNproofs}.
Comparing Props.~\ref{prop:UESIC}~and~\ref{prop:UESICn}, it is obvious
that the inequality in Prop.~\ref{prop:UESICn} is asymptotically tight as
$n\to\infty$.
More precisely, since $P_{c}^\textnormal{SIC}\geq P_{c,n}^\textnormal{SIC}$,
we have
\begin{equation*}
	\sum_{k=1}^{n} (1-\eta)^{k-1} \eta p_k\leq P_{c,n}^\textnormal{SIC}\leq\sum_{k=1}^{\infty} (1-\eta)^{k-1} \eta p_k,
\end{equation*}
and the difference between the upper and lower bound decays (at least) exponentially with $n$.
Thus, the lower bound in Prop.~\ref{prop:UESICn} converges to the true value
at least exponentially fast with $n$.

Combining Props.~\ref{prop:UESIC}~and~\ref{prop:UESICn} with the results
given before, we can estimate the performance gain
of SIC in the HCN downlink.
In the following, we focus on two different scenarios to
analyze the performance of finite SIC capability.

\subsubsection{The High SIR Case}

First, we focus on the SMUD regime ($\theta\geq 1$).
Thanks to Prop.~\ref{prop:altbounds_pk},
this case has extra tractability since $\P(\xi_k^{-1}>\theta I_k)$
can be expressed in closed-form.
Thus, by applying Prop.~\ref{prop:UESICn},
we obtain a set of upper bounds on the coverage probability with finite SIC capability
\begin{equation}
	P_{c,n}^\textnormal{SIC}\leq \frac{\eta}{1-\eta}\sum_{k=1}^n 
	\frac{1}{\Gamma(1+k\beta)}\left(\frac{1-\eta}{\theta^{\frac{\beta}{2}(k+1)}\Gamma(1-\beta)}\right)^k.
\label{equ:Pcn_UB}
\end{equation}
For infinite SIC capability, by the same procedure, a closed-form upper bound on the coverage probability
can also be obtained when $\alpha = 4$ ($\beta = \frac{1}{2}$)
\begin{align}
	P_c^{\textnormal{SIC}}	&=\sum_{k=1}^\infty (1-\eta)^{k-1} \eta p_k			\notag \\
						&\leq \sum_{k=1}^\infty (1-\eta)^{k-1} \eta \frac{(\pi\theta)^{-k/2}}{\Gamma(k/2+1)}		\label{equ:Pcn_UB_n=n} \\
						&= \frac{\eta}{1-\eta}\left(\exp\Big( \frac{(1-\eta)^2}{\pi\theta}			  \Big)\left(1+\operatorname{erf}\Big(\frac{1-\eta}{\sqrt{\pi\theta}}	\Big)\right)-1\right).
						\label{equ:Pcn_UB_n=infty}
\end{align}

\begin{figure}[t]
\centering
\psfrag{aaa}{\raisebox{-2pt}{\hspace{-10pt}$\theta=0$\;dB}}
\psfrag{bbb}{$\theta=2$\;dB}
\includegraphics[width=0.5\linewidth]{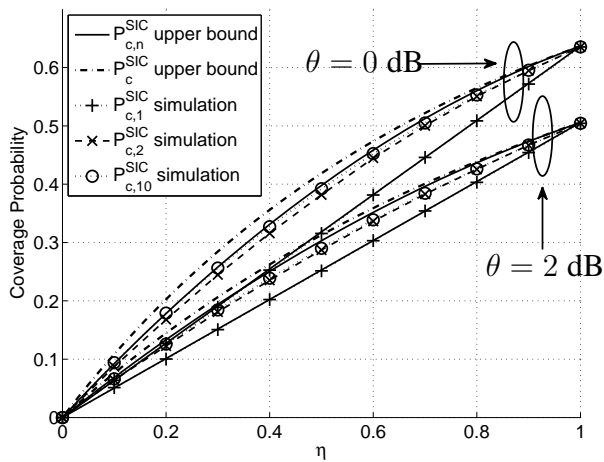}%
\caption{Comparison between the upper bound on the coverage probabilities 
and the simulated coverage probability of HCNs with different levels
of SIC capability when $\alpha=4$.
The upper bounds on $P_{c,n}^\textnormal{SIC}$ is calculated according
to (\ref{equ:Pcn_UB}) for $n=1,2$ (coverage probability is higher for larger $n$; note that when $n=1$ the bound is tight). 
The upper bound on $P_{c}^\textnormal{SIC}$ is calculated by (\ref{equ:Pcn_UB_n=infty}).
The simulated value of $P_{c,n}^\textnormal{SIC}$
is plotted for $n=1,2,10$.
When $\theta=2$\;dB, the curves for $n=2$ and $n=10$ almost completely overlap.}
\label{fig:FemtoSICCoverProbtheta1alpha4}
\end{figure}

Fig.~\ref{fig:FemtoSICCoverProbtheta1alpha4} plots the coverage probability with
different levels of SIC capability as a function of $\eta$ for $\theta=$0\;dB and 2\;dB.
Here, we plot the upper bounds on $P_{c,n}^\textnormal{SIC}$
according to (\ref{equ:Pcn_UB}) for $n=1,2$,
the upper bound on $P_{c}^\textnormal{SIC}$ according to (\ref{equ:Pcn_UB_n=infty}),
and simulated value of $P_{c,n}^\textnormal{SIC}$ for $n=1,2,10$.
The problem of $n=1$ is already studied in \cite{net:Dhillon12jsac}.

Taking $n=1$ and $\beta={1}/{2}$ in (\ref{equ:Pcn_UB}) and comparing it with (\ref{equ:Pcanalytical})
shows that the upper bound in (\ref{equ:Pcn_UB}) is tight for $n=1$.
This explains why the lowest solid lines (upper bound on $P_{c,1}^\textnormal{SIC}$)
and the lowest dashed lines (simulated $P_{c,1}^\textnormal{SIC}$) in Fig.~\ref{fig:FemtoSICCoverProbtheta1alpha4} overlap.

Fig.~\ref{fig:FemtoSICCoverProbtheta1alpha4} shows that
$P_{c,n}^\textnormal{SIC}-P_{c,1}^\textnormal{SIC}$, the absolute
coverage probability gain of SIC, is much larger when $\eta$ is close to ${1}/{2}$
than when $\eta$ is close to $0$ or $1$.
This phenomenon can be observed within a much wider range of system parameters.
Intuitively, this observation can be explained as follows:
On the one hand, when $\eta\to 1$, most of the BSs in the network
are accessible. Thus, SIC will not significantly improve the coverage probability, \ie there is \emph{no one to cancel}.
On the other hand, when $\eta \to 0$, most of the BSs in the network are non-accessible.
In this case, UE coverage can only be significantly improved if many
BSs are expected to be successively canceled.
As is shown in Section~\ref{sec:EN}, the number of BSs that can be successively decoded
is fundamentally limited by the choice of $\theta$,
and in this particular case ($\theta\geq 1$),
very few, if any, non-accessible BSs are expected to be canceled,
leaving very little space for SIC to improve the coverage probability, \ie the UE is \emph{unable to cancel}.

Moreover, it is worth noting that with $\theta\geq 1$ and $\beta=1/2$,
most of the gain of SIC is achieved
by the ability of canceling only a single non-accessible BS.
This is consistent with observations reported in \cite{net:Weber07tit}
where a different model for SIC is used and the transmission capacity is used as the metric.
The fundamental reason of this observation can be explained by Prop.~\ref{prop:UESICn}.
The difference in coverage probability between infinite SIC capability
and the capability of canceling $n-1$ UEs is $\sum_{k=n+1}^{\infty} (1-\eta)^{k-1} \eta p_k$,
which, due to the super-exponential decay of $p_k$ (Prop.~\ref{prop:altbounds_pk}),
decays super-exponentially with $n$.
Thus, most of the additional coverage probability comes from canceling a small number of non-accessible BSs.
Since $\theta$ affects the rate at which $p_k$ decays,
we can expect that the ability of successively decoding more than one non-accessible BS
becomes even less useful for larger $\theta$,
which is also demonstrated in Fig.~\ref{fig:FemtoSICCoverProbtheta1alpha4}.
This observation also implies that when $\beta=1/2$, $P^\textnormal{SIC}_c\approx P^\textnormal{SIC}_{c,2}\approx \frac{\eta}{\pi}(\frac{2}{\sqrt{\theta}}+\frac{1-\eta}{\theta}) $ thanks to \eqref{equ:Pcn_UB_n=n}.
When $\theta=0\;$dB, this approximation coincides with the upper bound on $P^\textnormal{SIC}_{c,2}$ plotted
in Fig.~\ref{fig:FemtoSICCoverProbtheta1alpha4}. Thus, its tightness can be observed by comparing 
the bound with simulated $P^\textnormal{SIC}_{c,10}$ in the case $\theta=0\;$dB.

Of course, with the same logic and analytical bounds, \emph{e.g.,}
the one in Prop.~\ref{prop:pk_hr_lb} or the one in Prop.~\ref{prop:pk_lr_lb},
we would expect that the ability to successively decode a large number of
BSs does help if $\beta\to 0$ and/or $\theta\to 0$.
$\beta\to 0$ could happen if the path loss exponent $\alpha$ is very large
and/or the BSs are clustered around the receiver and/or
the network dimension is low (\emph{e.g.} for vehicular networks, it is reasonable to take $d=1$).
$\theta\to 0$ happens when very low-rate codes are used.

\subsubsection{Other Realistic Cases}

Since the different values of $\theta$ and $\beta$ can result in
different usefulness of the finite SIC capability at the HCN downlink,
it is worthwhile to discuss the most realistic parameter choices in contemporary
systems.

The exact values of $\theta$ and $\beta$ depends on many facts including
modulation and coding schemes, receiver sensitivity, BS densities and propagation environment.
However, in practical OFDM-type HCNs (\emph{e.g.,} LTE and 802.11 networks), 
the SIR threshold $\theta$ is typically larger than $-3$\;dB and often more than $0$\;dB \cite{DhiGanC2011a}\footnote{The small $\theta$ regime is more applicable to wide-band systems, \emph{e.g.,} CDMA or UWB systems.}.
For the indoor propagation, $\alpha$ is typically between 3 and 4.
Therefore, the system parameters used in the high SIR case
(Fig.~\ref{fig:FemtoSICCoverProbtheta1alpha4})
are already reasonably realistic.

\begin{figure}[t]
\centering
\psfrag{aaaa}{\raisebox{-2pt}{\hspace{-5pt}$\alpha=3.7$}}
\psfrag{bbbb}{$\alpha=3.5$}
\psfrag{cccc}{$\alpha=3.3$}
\includegraphics[width=0.5\linewidth]{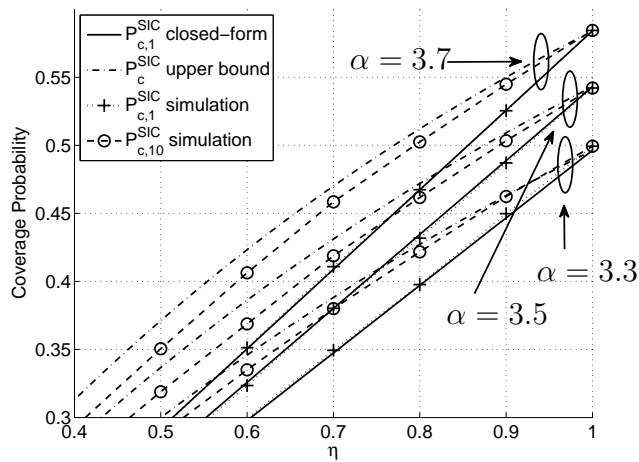}%
\caption{Comparison between the upper bound on the coverage probabilities 
and the simulated coverage probability of HCNs with different path loss exponents $\alpha$ when $\theta=0$\;dB.
The $P_{c,1}^\textnormal{SIC}$ is calculated by (\ref{equ:Pcnubalpha<4theta=1}) (note that when $n=1$ the bound is tight, \ie ``$\leq$" becomes ``$=$"). 
The upper bound on $P_{c}^\textnormal{SIC}$ is given by (\ref{equ:Pcubalpha<4theta=1}).
}
\label{fig:FemtoSICCoverProb_RealSim-many_alphas}
\end{figure}

To have a closer look at the impact of $\alpha$,
we fix $\theta=1$. Then, \eqref{equ:Pcn_UB} can be simplified as

\begin{equation}
	P_{c,n}^\textnormal{SIC}\leq \frac{\eta}{1-\eta}\sum_{k=1}^n 
	\frac{1}{\Gamma(1+k\beta)}\left(\frac{1-\eta}{\Gamma(1-\beta)}\right)^k.
	\label{equ:Pcnubalpha<4theta=1}
\end{equation}
which in the case of $n\to \infty$ gives an upper bound on the coverage probabilty with
infinite SIC capability,
\begin{equation}
	P_{c}^\textnormal{SIC}\leq \frac{\eta}{1-\eta} \left(\mathbf{E}_{\beta,1}\left(\frac{1-\eta}{\Gamma(1-\beta)}\right) -1 \right),
	\label{equ:Pcubalpha<4theta=1}
\end{equation}
where $\mathbf{E}_{a,b}(z)=\sum_{k=0}^\infty \frac{z^k}{\Gamma(ak+b)}$ is the Mittag-Leffler function.
Without resorting to the Mittag-Leffler function, one could also approximate $P_{c}^\textnormal{SIC}$
using \eqref{equ:Pcnubalpha<4theta=1} for small $n$. This is justified by the super-exponential decay
of $\frac{1}{\Gamma(1+k\beta)}\left(\frac{1-\eta}{\Gamma(1-\beta)}\right)^k$.
For example, when $\beta=1/2$, using second order approximation we get $P_{c}^\textnormal{SIC}\approx (3-\eta)\eta/\pi$.

Fig.~\ref{fig:FemtoSICCoverProb_RealSim-many_alphas}
compares the coverage probabilities with different levels of SIC capability
for different path loss exponents $\alpha$ when $\theta=1$.
As expected, as $\alpha$ decreases, both the coverage
probability and the gain of additional SIC capability decrease.
The former is due to the fact that with a smaller $\alpha$
the far BSs contribute more to the interference.
The latter can be explained by the fact that
when $\alpha$ is smaller,
the received power from different BSs are more comparable,
leaving less structure in the received signal that can be exploited
by SIC.

Similarly, we can apply the bounds in (\ref{equ:Pcnubalpha<4theta=1})
and (\ref{equ:Pcubalpha<4theta=1}) to even smaller $\alpha$ which
may apply to outdoor environments, and conceivably the gain of
SIC will becomes even more marginal.
Therefore, {\em SIC is more useful in an indoor environment}.

\begin{figure}[t]
\centering
\psfrag{aa}{$\eta=0.9$}
\psfrag{bb}{$\eta=0.6$}
\psfrag{cc}{$\eta=0.3$}
\includegraphics[width=0.5\linewidth]{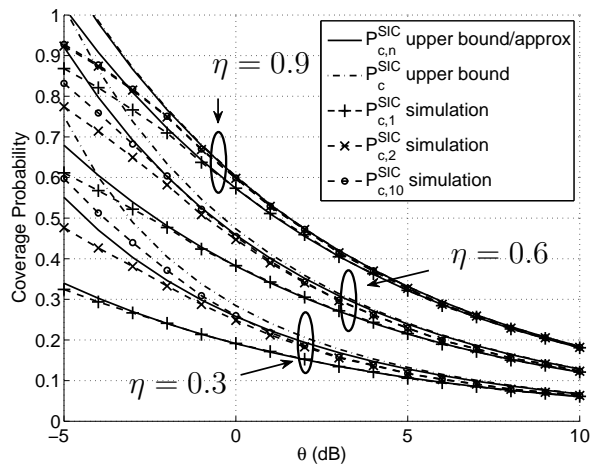}%
\caption{Coverage probability of HCNs with SIR threshold $\theta\geq -5$\;dB with $\alpha=4$.
The solid lines are calculated for $n=1,2$ according to (\ref{equ:Pcnubalpha<4theta=1})
(the lines are higher for larger $n$),
which are an upper bounds on $P_{c,n}^\textnormal{SIC}$ when $\theta\geq 0$\;dB.
For $\theta\leq 0$, these lines should be considered as approximations.
The upper bound on $P_{c}^\textnormal{SIC}$ is calculated by (\ref{equ:Pcubalpha<4theta=1}).
The simulated value of $P_{c,n}^\textnormal{SIC}$ is plotted for $n=1,2,10$.
For $\eta \leq 0.9$, the curves for $n=2$ and $n=10$ almost completely overlap throught the simulated SIR range.}
\label{fig:FemtoSICCoverProbwrt_thetaalpha4}
\end{figure}

Generally speaking, accurately estimating $P_{c,n}^\textnormal{SIC}$ is more difficult when $\theta<1$.
One of the reasons is that the upper bound in Thm.~\ref{thm:Pxik>thetaIk} becomes increasingly loose as $\theta$ decreases.
However, within the realistic parameters, \emph{i.e.,} $\theta>-3$\;dB,
the values calculated by (\ref{equ:Pcnubalpha<4theta=1})\footnote{
(\ref{equ:Pcnubalpha<4theta=1}) can only be considered as an approximation on $P_{c,n}^\textnormal{SIC}$
when $\theta<1$
since Prop.~\ref{prop:UESICn} only gives a lower bound in this regime.}
and (\ref{equ:Pcubalpha<4theta=1}) are still
informative as is shown in Fig.~\ref{fig:FemtoSICCoverProbwrt_thetaalpha4}.
This figure shows the coverage probability as a function of $\theta\geq -5$\;dB
for $\eta=0.3,0.6,0.9$.
We found that most of the conclusions we drew for $\theta\geq 1$
still hold when $\theta\geq -5$\;dB.
For example, we can still see that most of the gain of SIC comes from canceling a
single interferer and that the gain is larger when $\eta$ is close to $0.5$.

Quantitatively, we found that when $\eta$ is relatively small 
($\eta=0.3,0.6$) the analytical results still
track the results obtained by simulation closely for $\theta>-3$\;dB.
The analytical results are less precise when $\eta$ is large.
However, large $\eta$ characterizes a regime where most of the BSs are accessible.
In this case, it is conceivable that SIC is often unnecessary, which can be verified by either
the simulation results or the analytical results in Fig.~\ref{fig:FemtoSICCoverProbwrt_thetaalpha4}.
Therefore, overall, the analytical results generates enough quantitative insights
for the most interesting set of parameters.

\section{Conclusions\label{sec:conclu}}

Using a unified PLPF-based framework,
this paper analyzes the performance of SIC in
$d$-dimensional fading networks with power law density functions.
We show that the probability of successively
decoding at least $k$ users decays super-exponentially with $k^2$ if high-rate codes are used,
and that it decays especially fast under small path loss exponent in high dimensional networks,
which suggests the marginal gain of adding more SIC capability
diminishes very fast.
On the other hand, SIC is shown to be especially beneficial if very low-rate codes are used,
the active transmitters are clustered around the receiver,
or the dimensionality of the network is low, \eg $d=1$.

Since SIC can be considered not only as an interference mitigation technique but also as a
multiple packet reception (MPR) scheme,
we also investigate the performance gain of SIC in terms of aggregate throughput at the receiver,
counting information rate from all the decodable transmitters.
We observe that, in interference-limited networks, the aggregate throughput (or, sum rate)
is a monotonically decreasing function of the per-user information rate
and the asymptotic sum rate is $\frac{1}{\beta}-1$ as
the per-user information rate goes to $0$,
where $\beta=\frac{b+d}{\alpha}$, $\alpha$ is the path loss exponent and
$b$ determines the network geometry (clustering).
Since $b$ can be manipulated by distance-dependent access control or power control\cite{net:mh12},
the result shows that properly designed MAC or power control schemes
can significantly increase the network performance when combined with
SIC and low-rate codes (\eg in CDMA or ultra wide band (UWB) systems).

On the other hand, in noisy
networks, there exists at least one positive optimal
per-user rate which maximizes the aggregate throughput.
Moreover, different from interference-limited networks
where fading does not affect the performance of SIC \cite{Zhang12globecom},
we prove fading to be harmful in noisy networks. 
This suggests communication schemes that eliminate (average out)
the channel randomness
are desirable in noisy networks with SIC capability.

By a simple example, we demonstrate how the technical results in this paper
can be used to generate insights in designing heterogeneous cellular networks (HCNs) with SIC
capability at the UE side.
The results suggests that the choice of code rate can significantly impact the
usefulness of the capability of successively cancel a large number of interferers
in the downlink of HCNs.
In particular, SIC in combination with low rate codes can boost the coverage probability of the
HCNs to a large extent.
However, in terms of average non-outage throughput, the usefulness of SIC is relatively marginal.
Plugging in some realistic parameters of contemporary narrow band cellular systems,
we observe that, in uniform 2-d HCNs,
most of the gain of SIC comes from canceling a single interferer at the UE side.

\appendices

\section{Proof of Lemma~\ref{lem:d-dim} \label{app:d-dim}}

\begin{IEEEproof}
By Fact~\ref{fact:SPLPF}, $p_k$ can be evaluated by considering $\Xi_\beta$.
In particular, if we define the event $A_i=\{ \xi_i^{-1}>\theta I_{i}\}$,
the probability of successively decoding at least $k$ users
can be written as $%
	p_k=\P(\bigcap_{i=1}^{k}A_i).
$%

Defining $B_i\triangleq\{\xi_i^{-1}>(1+\theta) \xi_{i+1}^{-1}\}$,
we first show $\left( \bigcap_{i=1}^{k-1}B_i\cap A_k \right)\subset \bigcap_{i=1}^{k}A_i$ by induction. Consider the following statement:
\begin{equation}
\left( \bigcap_{i=n}^{k-1}B_i\cap A_k \right)\subset \bigcap_{i=n}^{k}A_i, \; n\leq k-1.
\label{equ:BnAk}
\end{equation}
\eqref{equ:BnAk} is true for $n=k-1$, since, for all $\omega\in B_{k-1} \cap A_k$, we have
\begin{equation*}
	\xi^{-1}_{k-1}(\omega)\stackrel{\text{\scriptsize (a)}}{>}
	\xi^{-1}_{k}(\omega)+\theta \xi^{-1}_{k}(\omega)\stackrel{\text{\scriptsize (b)}}{>}
	\theta I_{{k}}(\omega)+\theta \xi^{-1}_{k}(\omega)
	=	\theta I_{k}(\omega),
\end{equation*}
where (a) is due to $\omega\in B_{k-1}$, and (b) is due to $\omega\in A_k$.
In other words, the fact that $\omega\in B_{k-1} \cap A_k$ suggests $\omega \in A_{k-1} \cap A_{k}$
proves \eqref{equ:BnAk} for $n=k-1$.

Then, assuming \eqref{equ:BnAk} is true for $n\geq 2$, we can similarly show
that it is also true for $n-1$ by (again) considering an (arbitrary) realization of the PLPF
$\omega\in\bigcap_{i=n-1}^{k-1}B_i\cap A_k$. Since
\begin{equation*}
	\xi^{-1}_{n-1}(\omega)\stackrel{\text{\scriptsize (c)}}{>}
	\xi^{-1}_{n}(\omega)+\theta \xi^{-1}_{n}(\omega)\stackrel{\text{\scriptsize (d)}}{>}
	\theta I_{{n}}(\omega)+\theta \xi^{-1}_{n}(\omega)
	=	\theta I_{n}(\omega),
\end{equation*}
where (c) is due to $\omega\in B_{n-1}$, and (d) is due to $\omega \in \bigcap_{i=n}^{k-1}B_i$
and the assumption that \eqref{equ:BnAk} holds for $n$,
we have $\omega\in A_n$.
This proves \eqref{equ:BnAk} to be true for $n-1$ since $\omega\in\bigcap_{i=n}^{k-1}B_i\cap A_k
\subset \bigcap_{i=n}^{k}A_k $ by assumption.
Then, by induction, $\eqref{equ:BnAk}$ is shown to be true for $n=1$,
and thus
\begin{align}
	p_k	&\geq \P\left( \bigcap_{i=1}^{k-1}B_i\cap A_k \right)		
			=\E_{\xi_k}\left[	\P\left( \bigcap_{i=1}^{k-1}B_i\cap A_k\mid \xi_k \right) 	\right]	\notag\\
			&=\E_{\xi_k}\left[	\P\left( \bigcap_{i=1}^{k-1}B_i\right) \P\left( A_k \right) \mid \xi_k	\right], \label{equ:PBA}
\end{align}
where the last equality is because of the conditional independence between $B_i,\;\forall i\in[k-1]$
and $A_k$ given $\xi_k$.
Here, by definition, $\P\left( \bigcap_{i=1}^{k-1}B_i\right)=
	\P\left(\frac{\xi_i}{\xi_{i+1}}<(1+\theta)^{-1},\forall i<k \right)$.

Due to the Poisson property, conditioned on $\xi_{k},\;k\geq 2$,\footnote{
$\xi_k=x$ implies having $k-1$ points
on the interval $[0,x)$.}
we have $\frac{\xi_i}{\xi_{k}} \stackrel{\text{\scriptsize d}}{=} X_{i:k-1},\; \forall 1\leq i\leq k-1$, where $\stackrel{\text{\scriptsize d}}{=}$ means equality in distribution, $X$ is a random variable with cdf $F(x)=x^{\beta}\mathsf{1}_{[0,1]}(x)$,
and $X_{i:k-1}$ is the $i$-th order statistics of $k-1$ iid random variables 
with the distribution of $X$, \emph{i.e.,} the $i$-th smallest one
among $k-1$ iid random variables with the distribution of $X$.

Since $X^{\beta}\sim\textnormal{Uniform}(0,1)$, we can apply a result from the order statistics
of uniform random variables \cite{bk:OrderStatistics}.
In particular, if $U\sim\textnormal{Uniform}(0,1)$, then
$\left(\frac{U_{i:k-1}}{U_{i+1:k-1}}\right)^i \sim \textnormal{Uniform}(0,1)$
and $\left(\frac{U_{i:k-1}}{U_{i+1:k-1}}\right)^i$ is iid for all $1\leq i\leq k-2$.
Therefore, 
\begin{align}
\phantom{=}\; \P\left(\frac{\xi_i}{\xi_{i+1}}<(1+\theta)^{-1},\forall i<k\mid \xi_k\right)	
=\prod_{i=1}^{k-1} \P(U<(1+\theta)^{-i\beta})
	=(1+\theta)^{-\frac{\beta}{2}k(k-1)}, \label{equ:Xibeta}
\end{align}
where the last inequality is due to $\left(\frac{X_{i:k-1}}{X_{i+1:k-1}}\right)^{i\beta}\stackrel{\text{\scriptsize d}}{=} U, \;\forall i\in[k-2]$.
The lower bound is thus proved by combining (\ref{equ:PBA}) and (\ref{equ:Xibeta}).

Defining $\hat{B}_i = \{\xi_i^{-1}>\theta \xi_{i+1}^{-1}\}$
in the place of $B_i$, we can derive the upper bound in a very similar way.
\end{IEEEproof}

\section{Proofs of Lemma~\ref{lem:boundconditionalpk},~Prop.~\ref{prop:pk_lr_lb}, and Lemma~\ref{lem:upboundconditionalpk2}
\label{app:proofsofLHbounds}}

\begin{IEEEproof}[Proof of Lemma~\ref{lem:boundconditionalpk}]
In order to establish the lower bound,
we first calculate the mean of the interference $I_{k}$
conditioned on $\xi_k=\rho$, and then derive the bound based on the Markov inequality.
Denoting $I_k \mid \{\xi_k=\rho \}$ as $I_\rho$,
we can calculate the conditional mean interference by Campbell's Thm. \cite{net:mh12}
\begin{equation*}
	\E[I_\rho]=\E\bigg[\sum_{x\in\Xi\cap [\rho,\infty)}x^{-1}\bigg]
	=\int_\rho^\infty x^{-1} \Lambda(\d x)=\frac{a\beta}{1-\beta}\rho^{\beta-1}.
\end{equation*}
Thus, by the Markov inequality,
\begin{align*}
\P(\xi_k^{-1}>\theta I_{k}\mid\xi_k=\rho)= \P(\rho^{-1}>\theta I_\rho) \geq 1-\theta \rho\E[I_\rho].
\end{align*}
The lower bound can be refined as
$\left[1-\theta \rho\E[I_\rho]\right]^{+}$, where $[\cdot]^+=\max\{0,\cdot\}$.
Deconditioning over the distribution of $\xi_k$ (given by Lemma~\ref{lem:kNN}) 
yields the stated lower bound.
\end{IEEEproof}

\begin{IEEEproof}[Proof of Prop.~\ref{prop:pk_lr_lb}]
Using Fact~\ref{fact:SPLPF}, we work with $\Xi_\beta=\{\xi_i\}$.
For all $n\in[k-1],~k<1/\theta+1$,
we have
\begin{align*}
&\P \left(	\left\{	\xi_n^{-1}>\frac{\theta I_{n}}{1-(n-1)\theta} 	\right\} 
								\cap \left\{\xi_i>\theta I_{i},\;n< i \leq k\right\}	\right)		\\
& \qquad \stackrel{\text{\scriptsize{(a)}}}{\geq}		
			\P \left(	\left\{	\xi_{n+1}^{-1}>\frac{\theta I_{n}}{1-(n-1)\theta} 	\right\} 										\cap \left\{\xi_i>\theta I_{i},\;n< i \leq k\right\}	\right)		\\
& \qquad \peq{b}			\P \left(	\left\{	\xi_{n+1}^{-1}>\frac{\theta I_{{n+1}}}{1-n\theta} 	\right\} 
								\cap \left\{\xi_i>\theta I_{i},\;n< i \leq k\right\}	\right)		\\
& \qquad \peq{c}						\P \left(	\left\{	\xi_{n+1}^{-1}>\frac{\theta I_{{n+1}}}{1-n\theta} 	\right\} 
								\cap \left\{\xi_i>\theta I_{i},\;n+1 < i \leq k\right\}	\right),
\end{align*}
where (a) is because of the ordering of $\Xi$,
(b) is due to $I_{{n}}=\xi_{n+1}^{-1} + I_{{n+1}}$,
and (c) is due to the fact that
$
\left\{	\xi_{n+1}^{-1}>\frac{\theta I_{{n+1}}}{1-n\theta}\right\}\subset \left\{	\xi_{n+1}^{-1}>\theta I_{{n+1}}\right\}.
$
Using the inequality above sequentially for $n=1,2,\cdots,k-1$ yields
\begin{equation*}
	p_k\geq \P\left(\xi_k^{-1}>\frac{\theta I_k}{1-(k-1)\theta}\right),
\end{equation*}
where a lower bound for the RHS is given by Lemma~\ref{lem:boundconditionalpk} (substituting $\theta$ with $\tilde{\theta}$).
\end{IEEEproof}

\begin{IEEEproof}[Proof of Lemma~\ref{lem:upboundconditionalpk2}]
For a non-fading 1-d network,
the Laplace transform of the aggregate interference from $[\rho,\infty)$ can be calculated by
the probability generating functional (PGFL) of the PPP\cite{net:Haenggi08book}.
Similarly, the Laplace transform of $I_\rho\triangleq I_k\mid \{\xi_k=\rho\}$ is
\begin{equation}
\begin{split}\label{equ:PGFL}
	\phantom{=}\;\;\L_{I_\rho}(s)	&=\exp\left(-\int_\rho^\infty (1-e^{-sr^{-1}}) \Lambda(\d r) \right)	\\	
									&=\exp\left(-\Big(s^\beta \int_0^{s \rho^{-1}} r^{-\beta} e^{r} \d r - \rho^\beta (1-e^{-s \rho^{-1}})\Big)\right),
\end{split}
\end{equation}
where $\Lambda(\cdot)$ is the intensity measure of the SPLPF $\Xi_\beta$ (see Def.~\ref{def:SPLPF}).

Let $H$ be an exponential random variable with unit mean and independent of PLPF $\Xi$.
We can relate $\P(\xi_k^{-1}>\theta I_{k})$ with $\L_{I_{k}}(s)$ as
\begin{align*}
\P(\xi_k^{-1}>\theta I_{k})
	&= e \P(H>1) \P(\xi_k^{-1}>\theta I_{k})	
	\peq{a} e \P(\xi_k^{-1}>\theta I_{k},H>1)		\\
	&\leq e \P(H \xi_k^{-1}>\theta I_{k})				
	\peq{b}e\E_{\xi_k}[\L_{I_{k}\mid \xi_k}(\theta \xi_k)]	
	\peq{c} \E_{\xi_k}\left[\exp\left(-[c\xi_k^\beta -1]^{+}\right)\right],
\end{align*}
where (a) is due to the independence between $H$ and $\Xi$,
(b) is due to the well-known relation between the Laplace transform of
the interference and the success probability over a link subject to Rayleigh fading \cite{net:Haenggi08book},
(c) makes use of the PGFL in (\ref{equ:PGFL}), taking into account the fact that $\P(\xi_k^{-1}>\theta I_{k})\leq 1$.
With the distribution of $\xi_k$ given by Lemma~\ref{lem:kNN},
the proposition is then proved by straightforward but tedious
manipulation.
\end{IEEEproof}

\section{Proof of Thm.~\ref{thm:Pxik>thetaIk}\label{app:Pxik>thetaIk}}

First, we introduce the following lemma which is
necessary in proving Thm.~\ref{thm:Pxik>thetaIk}.

\begin{lemma}[Unique Decodeable Set]
\label{lem:k-strongest}
Consider an arbitrary $k$-element index set ${\cal K}\subset \mathbb{N}$
and an increasingly ordered set $\Xi=\{\xi_i\}$.
$\xi_{i}^{-1} >\theta \sum_{j\not\in {\cal K}} \xi_j^{-1}$ always implies 
$\xi_{i}^{-1} >\theta \sum_{j >k} \xi_j^{-1},~\forall i\leq k$.
Moreover, if $\theta\geq 1$ and $\xi_{i}^{-1} >\theta \sum_{j\not\in {\cal K}} \xi_j^{-1}$,
then ${\cal K}=[k]$.
\end{lemma}

\begin{IEEEproof}
The first part of the lemma is obviously true when ${\cal K}=[k]$.
If not, for any $l\in{\cal K}\backslash [k]$, we have $\xi_i^{-1} > \xi_l,~\forall i\in[k]$ by the ordering of $\Xi$.
For the same reason, we have $\sum_{j\not\in {\cal K}} \xi_j^{-1}>\sum_{j\not\in [k]} \xi_j^{-1}$.
As $\xi_l^{-1}>\theta\sum_{j\not\in {\cal K}} \xi_j^{-1}$, we have $\xi_i^{-1} > \sum_{j\not\in [k]} \xi_j^{-1},~\forall i\in[k]$.

To show the second part, consider an arbitrary $l\in{\cal K}$.
Since all elements in $\Xi$ are positive and $\theta\geq 1$, $\xi_{l}^{-1} >\theta \sum_{j\not\in {\cal K}} \xi_j^{-1}$
implies $\xi_{l} < \xi_j,~\forall j\not\in {\cal K}$, and consequently ${\cal K}=[k]$.
\end{IEEEproof}

Lemma~\ref{lem:k-strongest} states a general property of infinite
countable subsets of the real numbers.
Consider the case of $k=1$.
The second part of Lemma~\ref{lem:k-strongest} shows that if $\theta\geq 1$,
there is at most one user ($\xi_1$) that can be decoded without the help of SIC,
and this is always true even after an arbitrary number of cancellations.
In other words, multiple packet reception (MPR) is not feasible through parallel decoding.
This is exactly the reason why $\theta\geq 1$ is defined as sequential multi-user decoding (SMUD) regime. 

With Lemma~\ref{lem:k-strongest}, we now
give the proof of Thm.~\ref{thm:Pxik>thetaIk}.

\begin{IEEEproof}[Proof of Thm.~\ref{thm:Pxik>thetaIk}]
Consider the SPLPF (which is essentially a 1-d PPP) $\Phi\subset\R^+$ with intensity measure $\Lambda([0,r])=r^\beta$.
For each element $x\in\Phi$ we introduce an iid mark $h_x$ with exponential distribution with unit mean.
Since the marks $h_x$ can be interpreted as an artificial fading random variable,
in the following, we will refer this marked process as a path loss process with induced fading (PLPIF) $\hat{\Phi}\subset \R^+\times \R^+$.\footnote{The purpose of the induced fading may not be clear at
the moment. In particular, since we have already seen that one of the purposes of constructing the PLPF is to
`eliminate' fading as an explicit source of randomness,
constructing a PLPIF may seem to be one step backwards.
However, this is not the case due to the following subtlety: the PLPF
incorporates the randomness from an \emph{arbitrary} distribution into a 1-d PPP,
while the PLPIF is designed to facilitate the analysis by considering a \emph{particular}
fading random variable, \ie a unit mean exponential random variable.
}
Similar as before, based on $\hat{\Phi}$, we can construct a PLPF $\Xi(\hat{\Phi})=\{\hat{\xi}_i\}$ 
by letting $\hat\xi_i=\frac{x}{h_{x}},~\forall x\in\Phi$,
where, without loss of generality, we assume the indices $i$ are introduced such that $\Xi(\hat{\Phi})$ is increasingly ordered.

By Cor.~\ref{cor:fadingdoesntmatter}, we see that $p_k(\Xi(\hat{\Phi}))=p_k(\Xi_\beta)$.
Using the same technique in the proof of Prop.~\ref{prop:scale-invariant}, we can easily show that
\begin{equation}
	\P(\xi_k>\theta I_k)=\P(\hat{\xi}_k>\theta \hat{I}_k),~\forall k\in\mathbb{N},
\label{equ:nohat=hat}
\end{equation}	
where
$\hat{I}_k=\sum_{i=k+1}^\infty \hat{\xi}_i^{-1}$.\footnote{Note that we do \emph{not} have
$\hat\xi_i = \frac{\xi_i}{h_{\xi_i}}$ in general. In fact, the ordering
of $\Phi$ will \emph{not} be used in the rest of the proof.}
Therefore, in the following, we focus on the PLPIF $\hat{\Phi}$.

First, considering a $k$-tuple of positive numbers
${\bf y}=(y_i)_{i=1}^k\in(\R^+)^k$, with a slight abuse of notation,
we say $ (y_i)_{i=1}^{k} \subset \Phi$ if and only if $y_i\in\Phi,~\forall i\in[k]$.
Conditioned on ${\bf y} \subset \Phi$, we denote the interference from the rest of the network
$\sum_{x \in\Phi \backslash {\bf y}} h_x x^{-1}$ as $I^{!{\bf y}}$.
Since $\{y_i,~i\in[k]\}$ is a set of Lebesgue measure zero, by Slivnyak's theorem, we have
${I^{!{\bf y}}} \stackrel{\textnormal d}{=} I = \sum_{x \in\Phi } h_x x^{-1}$.
Thus, 
\begin{align}
	\L_I^{!{\bf y}}(s)&\triangleq\E[\exp(-sI^{!\bf y})]=\L_I(s)=\exp\left(-\E_h\left( \int_0^\infty \left( 1-\exp(-shr^{-1}) dr^\beta \right) \right)\right)	\notag	\\
	&=\exp\left(-\frac{s^\beta}{\sinc\beta}\right),
	\label{equ:LaplaceKfold}
\end{align}
where $\sinc x = \frac{\sin (\pi x)}{\pi x}$ and the derivation exploits the fact that $h_x$ are iid exponential random variables
with unit mean.

Second, let $ \mathcal{\hat N}$ be the sample space of $\hat\Phi$ and consider the indicator function $\bar\chi_k: (\R^+\times\R^+)^k\times\mathcal{\hat N}\to \{0,1\}$ defined as follows
\begin{equation*}
\bar\chi_k\big((x_i, h_{x_i})_{i=1}^k, \hat\phi\big)=
	\left\{
		\begin{array}{ll}
		1,&\textnormal{if } h_{x_i}x_i^{-1}>\theta \sum_{y\in\phi\backslash\{x_j,~j\in[k]\}}h_y y^{-1},~\forall i\in[k] \\
		0,&\textnormal{otherwise},
		\end{array}
	\right.
\end{equation*}
where $\phi\subset \R^+$ is the ground pattern of the marked point pattern $\hat\phi$.
In words, $\bar\chi_k\big((x_i, h_{x_i})_{i=1}^k, \hat\phi\big)$ is one iff 
$k$ of the users in the network
$(x_i)_{i=1}^k$ all have received power larger than
$\theta$ times the interference from the rest of the network.
Then, for any $\hat\phi$ and $k\in \mathbb{N}$,
\begin{align}
{\sf 1}_{\{\hat\xi_k>\theta \hat I_k\}}(\hat\phi) =
{\sf 1}_{\{\hat\xi_i>\theta \hat I_k,~\forall i\in[k]\}}(\hat\phi) 
\pleq{a}
\frac{1}{k!}\sum_{ x_1,\ldots,x_k \in\phi}^{\neq}\bar\chi_k\big((x_i, h_{x_i})_{i=1}^k, \hat\phi\big),
\label{equ:overcount.k!}
\end{align}
where $\neq$ means $x_i \neq x_j,\;\forall i\neq j$
and (a) is due to the first part of Lemma~\ref{lem:k-strongest}.
Also, the second part of Lemma~\ref{lem:k-strongest} shows that
when $\theta\geq 1$ the equality in (a) holds.

Therefore, we have
\begin{align*}
\P(\hat{\xi_k}^{-1}>\theta\hat{I}_k)
	=\E[{\sf 1}_{\{\hat\xi_k>\theta \hat I_k\}}(\hat\Phi)]		
	&\pleq{b}\frac{1}{k!}\E\left[\sum_{ x_1,\ldots,x_k \in\Phi}^{\neq}\bar\chi_k\big((x_i, h_{x_i})_{i=1}^k, \hat\Phi\big)\right]		\\
	&=\frac{1}{k!}\E_\Phi \left[\sum_{ x_1,\ldots,x_k \in\Phi}^{\neq} \E\left[\bar\chi_k\big((x_i, h_{x_i})_{i=1}^k, \hat\Phi\big)\right]\right]	\\
	&\peq{c}\frac{1}{k!}\E_\Phi \left[\sum_{{\bf x}:\; x_1,\ldots,x_k \in\Phi}^{\neq} \L_{I}^{!{\bf x}}(\theta\sum_{i=1}^k x_i) \right]	\\
	&\peq{d}\frac{1}{k!}\int_{(\R^+)^k} \L_{I}^{!{\bf x}}(\theta\sum_{i=1}^k x_i) \Lambda^{(k)}(\d {\bf x}),
\end{align*}
where (b) is due to (\ref{equ:overcount.k!}) and the equality holds when $\theta\geq 1$,
(c) holds since $h_y$ are iid exponentially distributed with unit mean for all $y\in\Phi$,
and (d) is due to the definition of $\Lambda^{(k)}(\cdot)$, the $k$-th factorial moment measure of $\Phi$ \cite[Chapter 6]{net:mh12}.
Since $\Phi$ is a PPP with intensity function $\lambda([0,r])=r^\beta$, we have $\Lambda^{(k)}({\d \bf x})=\prod_{i\in [k]}\d x_i^\beta$. 
Applying (\ref{equ:nohat=hat}) and (\ref{equ:LaplaceKfold}), we have
\begin{equation*}
	\P(\xi_k^{-1}>\theta I_k)\leq \frac{1}{k!} 
	\int_{(\R^+)^k} \exp\left( -\frac{ \theta^\beta  }{\sinc \beta } \|{\bf x}\|_{\frac{1}{\beta}} \right)\d {\bf x}.
\end{equation*}
where $\|\cdot \|_p$ denotes the $L_p$ norm, and the equality holds when $\theta\geq 1$.
The integral on the RHS can be further simplified into
closed-form by using the general formulas in \cite[eqn. 4.635]{GradShteynRyzhik},
which completes the proof.
\end{IEEEproof}

\section{Proofs of Prop.~\ref{prop:ENub}, Lemma~\ref{lem:LxikIk}, and Prop.~\ref{prop:scale-monotonicity} \label{app:ENATPf}}

\begin{IEEEproof}[Proof of Prop.~\ref{prop:ENub}]
By Prop.~\ref{prop:pk_c_ub},
we have $\E[N]\leq \sum_{k=1}^\infty \Delta_2(k)$.
The proposition then follows by summing up the first $K-1$ terms of the infinite
series and upper bounding the residue part.
Specifically, we have
\begin{align*}
\sum_{k=K}^\infty \frac{\gamma(k,1/c)}{\Gamma(k)}
	&= e^{-1/c}\sum_{k=K}^\infty \sum_{j=0}^\infty \frac{(1/c)^{j+k}}{(j+k)!}
	\stackrel{\text{\scriptsize{(a)}}}{\leq} \frac{\exp(-1/c)}{\sqrt{2\pi}}	\sum_{k=K}^\infty \sum_{j=0}^\infty
			\frac{(e/c)^{j+k}}{(j+k)^{j+k+\frac{1}{2}}}		\\
	&\leq \frac{\exp(-1/c)}{\sqrt{2\pi}} \sum_{k=K}^\infty \frac{(e/c)^k}{K^k}
			\sum_{j=0}^\infty	\frac{(e/c)^j}{(j+K)^{j+\frac{1}{2}}},
\end{align*}
where (a) uses Stirling's approximation for $n!$, \emph{i.e.,}
$\sqrt{2\pi}n^{n+1/2}e^{-n}\leq n! \leq e n^{n+1/2}e^{-n}$.
Moreover, $\sum_{j=0}^\infty	\frac{(e/c)^j}{(j+K)^{j+\frac{1}{2}}}
\leq e+\sum_{j=1}^\infty	\frac{(e/c)^j}{j^{j+\frac{1}{2}}}$ since $K\geq 1$.
Using Stirling's approximation again on
$\sum_{j=0}^\infty	\frac{(e/c)^j}{j^{j+\frac{1}{2}}}$
yields
\begin{equation*}
\sum_{k=K}^\infty \frac{\gamma(k,1/c)}{\Gamma(k)}
	\leq \frac{\exp(K+1)}{\sqrt{2\pi}}\frac{(cK)^{1-K}}{cK-1}.
\end{equation*}

Furthermore, we have
\begin{equation*}
	\sum_{k=K}^\infty \frac{e}{(1+c)^k}\frac{\Gamma(k,1+1/c)}{\Gamma(k)}
	\leq \sum_{k=K}^\infty \frac{e}{(1+c)^k} = \frac{e}{c}(1+c)^{1-K},
\end{equation*}
which completes the proof.
\end{IEEEproof}

\begin{IEEEproof}[Proof of Lemma~\ref{lem:LxikIk}]
As in the proof of Lemma~\ref{lem:upboundconditionalpk2},
we consider $I_\rho\triangleq I_k\mid \{\xi_k=\rho\}$ and the Laplace transform of $I_\rho$
is given in (\ref{equ:PGFL}).
Then, considering another random variable $\rho I_\rho \triangleq \xi_k I_k\mid \{\xi_k=\rho\}$,
we have
\begin{equation}
	\L_{\rho I_\rho}(s)=\E[e^{-s \xi_k I_k} \mid \xi_k=\rho]=\L_{I_\rho}(s\rho)=\exp(-c(s)\rho^\beta),
\end{equation}
where $c(s)=s^\beta\gamma(1-\beta,s)-1+e^{-s}$.
Using the results in Lemma~\ref{lem:kNN}, we can calculate the Laplace transform
of $\xi_k I_k$,
\begin{align*}
	\L_{\xi_k I_k }(s) &= \E_{\xi_k}[\L_{\rho I_\rho}(s) \mid \xi_k=\rho]		
											= \int_0^{\infty}\frac{\beta x^{k\beta-1}}{\Gamma(k)}e^{-(1+c(s))x^{\beta}}
											\d x
											=\frac{1}{(1+c(s))^k}.
\end{align*}
\end{IEEEproof}

\begin{IEEEproof}[Proof of Prop.~\ref{prop:scale-monotonicity}] 
The proof is similar to that of Prop.~\ref{prop:scale-invariant}.
Consider the mapping, $f(x)=(a_1/a_2)^{1/\beta}x^\beta$.
Then, $f(\Xi)$ is a PPP on $\R^+$ with intensity measure $a_2 x^\beta$ over the set $[0,x]$ for all $x>0$. 
As before, let $\mathcal{N}$ be the sample space of $\Xi$, \emph{i.e.,}
the family of all countable subsets of $\R^+$.
Consider an indicator function $\chi_k^W(\phi):\mathcal{N}\to \{0,1\} ,\; k\in\mathbb{N}$  such that
\begin{equation*}
	\chi_k^W(\phi)=
	\left\{
		\begin{array}{ll}
		1,& \textnormal{if  }	\xi_i^{-1}>\theta (I_i+W),~\forall i\leq k	
		\\
		0,&	\textnormal{otherwise,}
		\end{array}
	\right.
\end{equation*}
where $\phi=\{\xi_i\}$ and $\xi_i<\xi_j,\;\forall i<j$.

Note that $\chi_k^W(\phi)\leq \chi_k^W(C\phi),~\forall C\in(0,1)$, where $C\phi=\{C\xi_i\}$.
To show that, assume that $\chi_k^W(\phi)=1$, 
\emph{i.e.,} $\xi_i^{-1}>\theta (\sum_{j=i+1}^{\infty}\xi_j^{-1}+W),~\forall i\leq k$,
which is equivalent to $(C\xi_i)^{-1}>\theta (\sum_{j=i+1}^{\infty}(C\xi_j)^{-1}+C^{-1}W),~\forall i\leq k$.
It follows that $\chi_k^W(C\phi)=1$ since $C^{-1}W>W$. 

Therefore, we have
\begin{equation*}
	p_k(\Xi)=\E [\chi_k^W(\Xi)]\stackrel{\text{(a)}}{\leq}\E [\chi_k\left(f(\Xi)\right)]\stackrel{\text{(b)}}{=}\E [\chi_k\left(\bar{\Xi}\right)]=p_k(\bar{\Xi}),
\end{equation*}
where (a) is due to $a_1<a_2$ and thus $(a_1/a_2)^{1/\beta}<1$ and (b) is because both $f(\Xi)$
and $\bar{\Xi}$ are PPPs on $\R^+$ with intensity measure $\mu([0,r])=a_2 r^\beta$.
\end{IEEEproof}

\section{Proofs of Props.~\ref{prop:UESIC} and~\ref{prop:UESICn} \label{app:HCNproofs}}

\begin{IEEEproof}[Proof of Prop.~\ref{prop:UESIC}]
Without loss of generality, we consider the marked PLPF corresponding to
the $K$-tier heterogeneous cellular BSs $\hat{\Xi}=\{(\xi_i,t_i)\}$,
where the index $i$ is introduced such that $\{\xi_i\}$ are increasingly ordered.
Let $\vartheta_k: \mathcal{N}\to \{0,1\} ,\; k\in\mathbb{N}$, be an indicator function 
such that
\begin{equation}
	\vartheta_k(\phi)\triangleq
		\left\{
			\begin{array}{ll}
				1, & \textnormal{if $\exists l\in\mathbb{N}$ s.t. $\chi_l(\phi)=1$ and $\xi_k^{-1}>\theta I_l^{!k}$}		\\
				0, & \textnormal{otherwise,}
			\end{array}
		\right.
		\label{equ:vartheta_k}
\end{equation}
where $\chi_k(\cdot)$ is defined in (\ref{equ:chi_k}).
Furthermore, we define a random variable $M=\min\{i:t_i=1\}$,
where $t_i$ is the mark of the $i$-th element in $\hat{\Xi}$.
Note that since, according to Lemma~\ref{lem:heteroPLPF},
$t_i$ are iid (also independent from $\Xi$),
$M$ is geometrically distributed with parameter $\eta$
and is independent of $\Xi$.
Then, it is easy to check with Def.~\ref{def:coverage}
that the coverage probability can be written as
\begin{equation*}
	P_c^\textnormal{SIC}=	\P(\vartheta_M(\Xi))
	=\E_M\left[	\P(\vartheta_M(\Xi)\mid M)	\right],
\end{equation*}
where the probability inside the expectation is the probability of decoding
the $M$-th strongest BS (with the help of SIC) conditioned on the fact that
this BS is the strongest accessible BS.

Moreover, we have $\vartheta_k(\cdot)=\chi_k(\cdot),~\forall k\in\mathbb{N}$.
To see this, we first notice that, by the definition of the two functions,
$\chi_k(\phi)=1\;\Rightarrow\;\vartheta_k(\phi)=1$.
Conversely, assuming $\vartheta_k(\phi)=1$, which by definition means $\exists l\in\mathbb{N}$ s.t. $\chi_l(\phi)=1$ and $\xi_k^{-1}>\theta I^{!k}_l$,
we immediately notice that $\chi_k(\phi)=1$ if $l\geq k$.
If $l<k$, we have $\xi_{l+1}^{-1}\geq\xi_k^{-1}
>\theta I_l^{!k}\geq\theta I_{l+1}$,
\emph{i.e.,} $\chi_{l+1}(\phi)=1$, which, by induction, leads to the
fact that $\chi_k(\phi)=1$. Since both $\chi_k(\cdot)$ and $\vartheta_k(\cdot)$
are indicator functions on the domain of all countable subsets of $\R^+$,
we have established the equivalence of the two functions.

Therefore, we have $P_c^\textnormal{SIC}=\E_M\left[	\P(\chi_M(\Xi)\mid M)	\right]=\E_M [p_M]$,
which completes the proof.
\end{IEEEproof}

\begin{IEEEproof}[Proof of Prop.~\ref{prop:UESICn}]
Similar to the definition of $\vartheta_k(\cdot)$ in (\ref{equ:vartheta_k}).
We define
\begin{equation}
	\vartheta_{n,k}(\phi)\triangleq
		\left\{
			\begin{array}{ll}
				1, & \textnormal{if $\exists l< n $ s.t. $\chi_l(\phi)=1$ and $\xi_k^{-1}>\theta I_l^{!k}$}		\\
				0, & \textnormal{otherwise.}
			\end{array}
		\right.
		\label{equ:vartheta_n,k}
\end{equation}
Then, we have
\begin{align*}
	P_{c,n}^\textnormal{SIC} &\peq{a} \E_M\left[	\P(\vartheta_{n,M}(\Xi)\mid M)	\right]			
													\peq{b} \sum_{k=1}^\infty \eta (1-\eta)^{k-1} \P(\vartheta_{n,k}(\Xi))		\\
													&\pgeq{c} \sum_{k=1}^n \eta (1-\eta)^{k-1} \P(\chi_k(\Xi))		
													\peq{d} \sum_{k=1}^n \eta (1-\eta)^{k-1} p_k.
\end{align*}
where (a) is due to Def.~\ref{def:coverageFinite},
(b) is due to the independence between the marks and the process $\Xi$
and (d) is due to the definition of $p_k$.
To show (c), we note that $\vartheta_{n,k}(\cdot)=\chi_k(\cdot)$ for all $k\leq n$,
which can be shown in a way analogous to the way we establish the equivalence between $\vartheta_k(\cdot)$
and $\chi_k(\cdot)$ in the proof of Prop.~\ref{prop:UESIC}.
In addition, when $\theta\geq 1$, for all $k>n>l$,
$\xi_k^{-1}<\theta\sum_{j\geq l+1}^{j\neq k}\xi_j^{-1}$ almost surely.
In other words, $\P(\vartheta_{n,k}(\cdot))=0$ for all $k>n$ and the equality in (c) is attained for $\theta\geq 1$.
\end{IEEEproof}


\bibliographystyle{IEEEtran}
\bibliography{IEEEabrv,../mynet}

\end{document}